\documentclass[aps,prd,nofootinbib,12pt]{revtex4}
\usepackage{epsfig}
\usepackage{graphicx}
\usepackage[font=small,skip=0pt,justification=raggedright]{caption}
\usepackage{amsmath,amssymb,mathrsfs}
\newcounter{fig}   \newcommand{\lbfig}[1]{\refstepcounter{fig}
\label{#1} }

\newcommand{\bea}{\begin{eqnarray}}
\newcommand{\eea}{\end{eqnarray}}
\newcommand{\be}{\begin{equation}}
\newcommand{\ee}{\end{equation}}

\newcommand{\re}[1]{(\ref{#1})}

\newcommand{\eqn}{\begin{eqnarray}}
\newcommand{\eqnx}{\end{eqnarray}}

\date{\today}

\begin{document}
\title{Q-chains  in the $U(1)$ gauged Friedberg-Lee-Sirlin model}
\author{V.~Loiko}
\affiliation{ Department of Theoretical Physics and Astrophysics,
Belarusian State University, Minsk 220004, Belarus}
\author{I.~Perapechka}
\affiliation{ Department of Theoretical Physics and Astrophysics,
Belarusian State University, Minsk 220004, Belarus}
\author{Ya.~Shnir}
\affiliation{BLTP, JINR, Dubna 141980, Moscow Region, Russia
}
\begin{abstract}
We construct static axially symmetric  multi-Q-ball configurations in the $U(1)$ gauged
two-component Fridberg-Lee-Sirlin model a flat spacetime. The solutions represent electromagnetically
bounded chains of stationary spinning charged Q-balls placed along the axis of symmetry.
 We discuss the properties of these configurations and exhibit their domain of existence.

\end{abstract}
\maketitle

\section{Introduction}
Q-balls represent spatially localized finite energy field configurations in flat space
\cite{Rosen,Friedberg:1976me,Coleman:1985ki}, they
may exist in a model with complex scalar field possessing an unbroken, continuous global symmetry
(for a review, see, e.g. \cite{Lee:1991ax,Shnir2018,Radu:2008pp}).
The Q-balls carry a Noether charge associated with this symmetry,
they are time-dependent non-topological solitons with a stationary oscillating  phase.
Configurations of this type may exist both in the models with a single complex scalar field and a suitable
non-renormalizable self-interaction potential \cite{Coleman:1985ki}, and in a two-component model with a
symmetry breaking potential
\cite{Friedberg:1976me}.

The Friedberg-Lee-Sirlin (FLS) model \cite{Friedberg:1976me} is a simple two-component
renormalizable scalar theory with fourth-order potential and a minimal interaction term.
In such a model the complex scalar field becomes massive due to the coupling between the
components, the real field has a finite vacuum expectation
value generated by the corresponding symmetry breaking potential.
Notably, stable non-topological soliton solutions of that
model also exist  in the limiting case of vanishing potential \cite{Levin:2010gp,Loiko:2018mhb} as the vacuum expectation value
of the real component remains finite.

Evidently, one can generalize the model considering gauged Q-balls with local $U(1)$ symmetry
\cite{Lee:1988ag,Lee:1991bn,Kusenko:1997vi,Anagnostopoulos:2001dh,Gulamov:2015fya,Gulamov:2013cra,Panin:2016ooo,Nugaev:2019vru,
Loginov:2020xoj,Loginov:2020lwg}. In the $U(1)$ gauged version
of the  FLS model \cite{Lee:1991bn},
the  Noether charge is associated with the harmonic time-dependency of the complex field. Further, the $U(1)$ gauged
Friedberg-Lee-Sirlin-Maxwell model can be considered as a truncated version of the  Witten's model
of superconducting cosmic strings with $U(1)\times U(1)$ local gauge invariance
\cite{Witten:1984eb,Loiko:2018mhb,Forgacs:2020vcy}.

Apart the fundamental spherically symmetric Q-balls, there are both radially and angularly
excited solutions \cite{Volkov:2002aj,Kleihaus:2005me,Kleihaus:2007vk}.
The radially excited Q-balls are still spherically symmetric, however
the scalar field possesses one or more nodes in radial direction.
Similar radially excited gauged Q-balls also exist in the $U(1)$ gauged model \cite{Loginov:2020lwg}.

The angularly excited axially symmetric Q-balls with non-zero angular
momentum possess an additional azimuthal phase factor of the spinning field
\cite{Volkov:2002aj,Kleihaus:2005me,Radu:2008pp}. In the $U(1)$ gauged theory gauged spinning Q-balls may induce a toroidal
magnetic field \cite{Shiromizu:1998eh,Loiko:2019gwk}.
Solutions of that type can be viewed as vortons, the finite energy
localized spinning loops stabilized by rotation
\cite{Witten:1984eb,Radu:2008pp,Davis:1988jp,Davis:1988ij,Garaud:2013iba}.

Typically, there are two branches of spinning gauged Q-balls. The lower in energy (electric) branch emerges
from the perturbative spectrum, as the angular frequency is decreasing below the mass threshold.
This branch of solutions terminates at some critical value of the frequency $\omega_{min}$,
here it bifurcates with the second,
higher in energy magnetic branch which extends backwards as the frequency increases.
The energy of the magnetic field of a circular vortex
rapidly grows along this branch, its strong magnetic field may destroy the superconductive phase in the interior
of the Q-ball \cite{Loiko:2019gwk}.

Further,
it was pointed out that there are two families of the spinning Q-balls with positive and negative parity, the
corresponding solutions are symmetric or antisymmetric with respect to reflections in the xy-plane \cite{Volkov:2002aj}.
The simplest parity-odd solution can be considered as a pair of Q-balls spinning in opposite phases.

It is known that gauged Q-balls may exist only for relatively small values of the gauge coupling
\cite{Lee:1988ag,Lee:1991bn,Anagnostopoulos:2001dh,Nugaev:2019vru}. Further, they are expected to become unstable
for large values of the Noether charge $Q$ because of the repulsive electric Coulomb force \cite{Lee:1988ag}.
A possibility that, to our best knowledge, has not been considered before, is that in the $U(1)$ gauged model
the electric repulsion can be balanced by the scalar and magnetic interactions of the axially symmetric spinning solitons.
This may open a way to construct a new type of solutions, which correspond to chains of gauged spinning Q-balls.
In these axially symmetric equilibrium configurations a number of constituents are located
symmetrically with respect to the origin along the symmetry axis, the
repulsive electric interaction is balanced by the scalar and magnetic forces between the
spinning gauged Q-balls providing zero net effect. Similar chain solutions are known to exist in various systems,
both for gravitating and flat space solitons, e.g. for
non-Abelian monopoles and dyons
\cite{Kleihaus:1999sx,Kleihaus:2000hx,Kleihaus:2003nj,Kleihaus:2003xz,Kleihaus:2004is,Teh:2004bq,
Paturyan:2004ps,Kleihaus:2004fh,Kleihaus:2005fs,Kunz:2006ex,Kunz:2007jw,Lim:2011ra,Teh:2014zea},
Skyrmions \cite{Krusch:2004uf,Shnir:2009ct,Shnir:2015aba}, and boson stars \cite{Herdeiro:2020kvf}.

A main objective of this Letter, which extends our previous consideration of the parity-even axially symmetric
spinning gauged Q-balls in the Friedberg-Lee-Sirlin-Maxwell theory \cite{Loiko:2019gwk},
is to examine this possibility.
We show that, indeed, there are new families of multi-component
axially-symmetric solutions of the  model, which represent chains of spinning gauged Q-balls, they
possess both a non-zero electric charge and
a magnetic field. We found such configurations with even number of $k$ constituents on the symmetry axis numerically
and determine their domains of existence.

\section{ The model and field equations}
We consider the four-dimensional $U(1)$-gauged two-component
Friedberg-Lee-Sirlin-Maxwell model, which describes a coupled system of the
real self-interacting scalar field $\psi$ and a complex scalar field $\phi$, minimally interacting
with the Abelian gauge field $A_\mu$. The corresponding Lagrangian density is
\be
L= -\frac14 F_{\mu\nu} F^{\mu\nu} + (\partial_\mu\psi)^2 + |D_\mu\phi|^2 - m^2 \psi^2|\phi|^2 - U(\psi) \, ,
\label{lag-fls}
\ee
where $D_\mu = \partial_\mu +igA_\mu$ denotes the covariant derivative.
The electromagnetic field strength tensor is $F_{\mu\nu}=\partial_\mu A_\nu-\partial_\nu A_\mu$
with electric components
$E_k=F_{k0}$ and magnetic components $B_k=\varepsilon_{kmn}F^{mn}$,
$g$ denotes the gauge coupling constant and $m$ is the scalar coupling constant, it defines the mass of the complex
component $\phi$. Without loss of generality we assume $g\ge 0$.

The symmetry breaking potential of the real scalar field $\psi$ is
\be
U(\psi)= \mu (1-\psi^2)^2 \, ,
\label{Pot}
\ee
thus, $\psi \to 1$ in the vacuum and the local $U(1)$ symmetry is broken in the interior of the Q-ball,
where the gauge field $A_\mu$
becomes massive. Evidently, the system \re{lag-fls} represents a generalization of the
Abelian Higgs model,  in other words,  the gauged Q-ball behaves like a
superconductor \cite{Lee:1988ag} with the field component $\psi$ playing a role of the order parameter.

On the other hand, the model \re{lag-fls} can be considered as a reduced
version of the
$U(1)\times U(1)$ gauged model of superconducting strings \cite{Witten:1984eb}.
Such a  theory supports stationary vortex rings stabilized by charge, current and
angular momentum, so called vortons \cite{Davis:1988jp,Davis:1988ij,Garaud:2013iba}.

The model \re{lag-fls} is invariant under the local $U(1)$ gauge
transformations of the fields. This symmetry results in the existence of the conserved
Noether current
\be
\label{Noether}
j_\mu = i(\phi D_\mu\phi^\ast-\phi^\ast D_\mu\phi) \, ,
\ee
with the corresponding charge $Q=\int{d^3x~ j^0}$. This electormagnetic
current is a source in the Maxwell equation
\be
\partial^\mu F_{\mu\nu}= g j_\nu
\label{eq-em}
\ee

Variation of the Lagrangian \re{lag-fls} with respect to the
fields $\psi$ and $\phi$ yields the field equations
\be
\label{scaleq}
\begin{split}
    \partial^\mu\partial_\mu \psi&=2\psi\left(m^2|\phi|^2+2\mu \left(1-\psi^2\right)\right),\\
    D^\mu D_\mu \phi&=m^2\psi^2\phi \, ,
\end{split}
\ee

We are interested in stationary spinning axially-symmetric solutions of the model \re{lag-fls}.
The corresponding parametrization of the scalar fields is
\be
\label{scalans}
\psi=X(r,\theta)\, , \qquad  \phi=Y(r,\theta)e^{i(\omega t+n\varphi)}\, ,
\ee
where $\omega$ is the angular frequency of the spinning complex field $\phi$, and
$n\in\mathbb{Z}$ is the azimuthal winding number. Without loss of generality we can consider
positive values of the angular frequency $\omega$.
Further, in the static gauge the electromagnetic potential
can be written as
\be
\label{Aans}
A_{\mu} dx^{\mu} =A_0(r,\theta)dt + A_\varphi(r,\theta) \sin\theta d\varphi
\ee
Substitution of the ansatz \re{scalans},\re{Aans} into the definition of the $U(1)$ charge $Q$ above
gives
\be
Q=\int d^3 x \left(gA_0+\omega \right) Y^2 \, ,
\ee

The stationary spinning axially symmetric configurations possess angular momentum
which is given by the  $T_\varphi^0$ component of the stress-energy tensor,
\be
J = \int d^3x~T_\varphi^0 = 4\pi \int\limits_0^\pi\int_0^\infty~r^2\sin\theta dr d\theta~\biggl\{
 (g A_0 + \omega ) (n + g A_\phi \sin\theta ) Y^2
+ J_{em}\biggr\} \, ,
\label{ang}
\ee
where we separated the angular momentum of the electromagnetic field
\be
J_{em}= \frac{1}{r^2} \partial_\theta A_0 \left(A_\phi\cos\theta + \sin\theta \partial_\theta A_\phi\right)
+\sin\theta  \partial_r A_\phi \partial_r A_0
\ee
The angular momentum of the spinning gauged Q-ball is quantized in the units of the electric charge of the configuration,
$J=nQ$ \cite{Radu:2008pp}.

The total energy of the system becomes
\be
\begin{split}
E&=2\pi \int\limits_0^\pi\int_0^\infty~r^2\sin\theta dr d\theta~\biggl\{
X_r^2+Y_r^2 + \frac{X_\theta^2}{r^2} + \frac{Y_\theta^2}{r^2} +
\frac{1}{r^2}\left(g A_\phi +\frac{n}{\sin \theta}\right)^2 Y^2\\
&+ (gA_0 + \omega)^2 Y^2 + \mu(1-X^2)^2 + m^2 X^2 Y^2 + E_{em}\biggr\} \, ,
\end{split}
\label{energy}
\ee
where $X_{r,\theta}\equiv\partial_{r,\theta}X$, $Y_{r,\theta}\equiv\partial_{r,\theta}Y$,
and the electromagnetic energy density is
$$
E_{em}= \frac12\left\{ (\partial_r A_0)^2 + \frac{1}{r^2}(\partial_\theta A_0)^2
+\frac{1}{r^2}(\partial_r A_\phi)^2 + \frac{1}{r^4\sin^2\theta}
\left[\partial_\theta(A_\phi \sin\theta)\right]^2\right\}
$$

The field equations resulting from the variation of the reduced action on the ansatz
\re{scalans},\re{Aans} are
\be
\begin{split}
&\left(\frac{\partial^2}{\partial r^2} +
\frac{2}{r}\frac{\partial}{\partial r} + \frac{1}{r^2} \frac{\partial^2}{\partial \theta^2}
+ \frac{\cos \theta}{r^2 \sin\theta}\frac{\partial}{\partial \theta}
+ 2 \mu^2 (1-X^2) - m^2 Y^2
 \right) X = 0\, ;\\
&
\left(\frac{\partial^2}{\partial r^2} +
\frac{2}{r}\frac{\partial}{\partial r} + \frac{1}{r^2} \frac{\partial^2}{\partial \theta^2}
+ \frac{\cos \theta}{r^2 \sin\theta}\frac{\partial}{\partial \theta}
-\frac{1}{r^2} \left(gA_\phi +\frac{n}{\sin\theta} \right)^2
+ (gA_0+\omega )^2 - m^2 X^2
 \right) Y = 0\, ;\\
&
\left(\frac{\partial^2}{\partial r^2} +
\frac{2}{r}\frac{\partial}{\partial r} + \frac{1}{r^2} \frac{\partial^2}{\partial \theta^2}
+ \frac{\cos \theta}{r^2 \sin\theta}\frac{\partial}{\partial \theta}
-\frac{1}{r^2\sin^2\theta} - 2 g^2 Y^2 \right) A_\phi = \frac{2n g}{\sin\theta} Y^2\, ;\\
&
\left(\frac{\partial^2}{\partial r^2} +
\frac{2}{r}\frac{\partial}{\partial r} + \frac{1}{r^2} \frac{\partial^2}{\partial \theta^2}
+ \frac{\cos \theta}{r^2 \sin\theta}\frac{\partial}{\partial \theta}
-2 g^2 Y^2\right) A_0 = 2g\omega Y^2 \, .
\end{split}
\label{eqs}
\ee
As usually, the last equation in the system \re{eqs} represents the Gauss law, a constraint imposed on
the system.

Note that, in the static gauge both the electric and magnetic components of the gauge potential are set to vanish as $r\to\infty$.
Further, in the limiting case $\omega^2\approx m^2$ the equations on the profile functions $X(r,\theta),Y(r,\theta)$ \re{eqs}
can be linearized. Then the asymptotic expansion of the system of equation \re{eqs} suggests that,
similar to the case of the spinning axially-symmetric Q-balls
\cite{Volkov:2002aj,Kleihaus:2005me,Kleihaus:2007vk,Radu:2008pp,Loiko:2018mhb},
the angular part of the asymptotic solutions for the complex field function $Y(r,\theta)$ is associated with the
real spherical harmonics $Y_{\ell n}(\theta,\varphi)$ which are proportional to the associated Legendre polynomials
$P_\ell^{n}(\cos \theta)$. Similarly, the angular part of the real field function $X(r,\theta)$ is proportional to the
Legendre polynomials $P_n(\cos \theta)$.

Thus, the field components of the axially symmetric spinning solutions of the
$U(1)$ gauged FLS model \re{lag-fls} may be either symmetric with respect to reflections in the equatorial
plane, $\theta \to \pi -\theta$, or  antisymmetric.
However, since the real component is approaching a non-zero vacuum value
$X(r,\theta)=1$ as $r\to \infty$, the field $\psi$ has to be parity-even.
For example, the spherical harmonic $Y^1_1(\theta,\varphi)$ yields
a parity-even component of the complex field, while the harmonic $Y^1_2(\theta,\varphi)$ corresponds to the parity-odd
component. The simplest spherically symmetric solution corresponds to the
spherical harmonic $Y^0_0$ for the complex component, and to the Legendre polynomial $P_0$, for the real component.

Recently, the existence of parity-even spinning
gauged Q-balls  was demonstrated \cite{Loiko:2019gwk}.
In this paper, we present strong numerical arguments that new mixed parity angularly excited solutions
of the nonlinear system of field equations \re{eqs} exist,
for those the function $Y(r,\theta)$ is parity-odd while the real component $X(r,\theta)$ is parity-even.

The system \re{eqs} represents a set of four coupled elliptic partial differential
equations with mixed derivatives. We solved this system numerically, by taking into account appropriate boundary conditions.
As usual, they follow from the condition of regularity of the fields at the origin and on the symmetry axis,
the requirements of symmetry, and the finiteness of the energy of the system. In particular we have to take into account that
in the static gauge the potential $A_\mu$
is vanishing at the spatial boundary, as the real field $X$ approaches its vacuum value.
Accordingly, we impose
\be
X\bigl.\bigr|_{r=\infty}=1  \, , \qquad Y\bigl.\bigr|_{r=\infty}=0
\, , \qquad  A_0\bigl.\bigr|_{r=\infty}=0
\, , \qquad A_\phi\bigl.\bigr|_{r=\infty}=0 \, .
\label{boundary_r}
\ee
Further, the restriction of regularity at the origin gives
\be
\partial_r X\bigl.\bigr|_{r=0}=0  \, , \qquad Y\bigl.\bigr|_{r=0}=0
\, , \qquad \partial_r A_0\bigl.\bigr|_{r=0}=0
\, , \qquad \partial_r A_\phi\bigl.\bigr|_{r=0}=0
\label{boundary_o}
\ee
Note that for spherically symmetric Q-ball the scalar field component $Y(r,\theta)$ has a finite value at the origin,
the corresponding boundary condition is $\partial_r Y\bigl.\bigr|_{r=0,\theta}=0$.

On the symmetry axis we have to impose
\be
\partial_\theta X\bigl.\bigr|_{\theta=0}=0  \, , \qquad Y\bigl.\bigr|_{\theta=0}=0 \, ,
\qquad  \partial_\theta A_0\bigl.\bigr|_{\theta=0}=0
\, , \qquad A_\phi\bigl.\bigr|_{\theta=0}=0\, ,
\label{boundary_ax}
\ee
and on the equatorial plane, for parity-odd solutions, we impose
\be
X\bigl.\bigr|_{\theta=\pi/2}=1  \, , \qquad Y\bigl.\bigr|_{\theta=\pi/2}=0 \, ,
\qquad  \partial_\theta A_0\bigl.\bigr|_{\theta=\pi/2}=0
\, , \qquad \partial_\theta A_\phi\bigl.\bigr|_{\theta=\pi/2}=0\, .
\label{boundary_pl}
\ee

We have solved the boundary value problem
for the coupled system of nonlinear partial differential equations \re{eqs}
with boundary conditions \re{boundary_r}-\re{boundary_pl} using
a six-order finite difference scheme. The numerical calculations are mainly performed on an equidistant grid
in spherical coordinates $r$ and $\theta$. Typical grids we used have sizes $121 \times 101$.
In our numerical scheme we map the infinite interval of the variable $r$ onto the compact radial coordinate
$x=\frac{r/r_0}{1+r/r_0}  \in [0:1]$. Here $r_0$ is a real scaling constant,
which is used to adjust the contraction of the grid.
Typically it is taken as $r_0 = 4 - 10$.
The underlying linear system is solved with the Intel MKL
PARDISO sparse direct solver \cite{pardiso} using the Newton-Raphson method. Calculations are performed with
the packages FIDISOL/CADSOL \cite{schoen} and CESDSOL\footnote{Complex Equations -- Simple
Domain partial differential equations SOLver is a C++ package
being developed by one of us (I.P.).
} library.
Estimated numerical errors are of order of $10^{-3}$. For convenience, in our numerical calculations we fix the value of the
mass parameter $\mu=0.25$, set the coupling
constant $m=1$ and restrict our consideration to angularly excited spinning gauged $n=1$ Q-balls.

\section{Numerical results}

\begin{figure}[h!]
\begin{center}
\includegraphics[height=4cm]{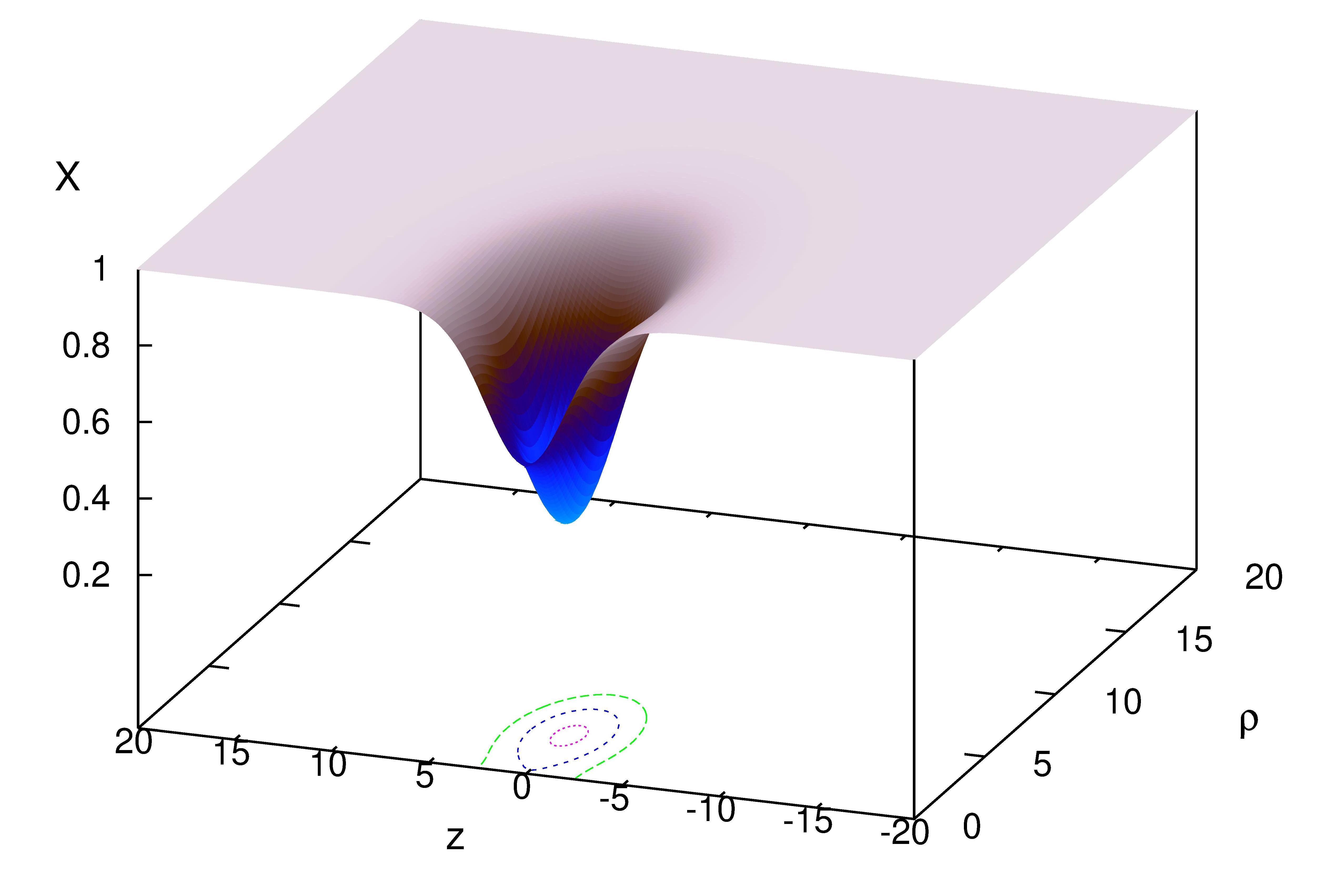}
\includegraphics[height=4cm]{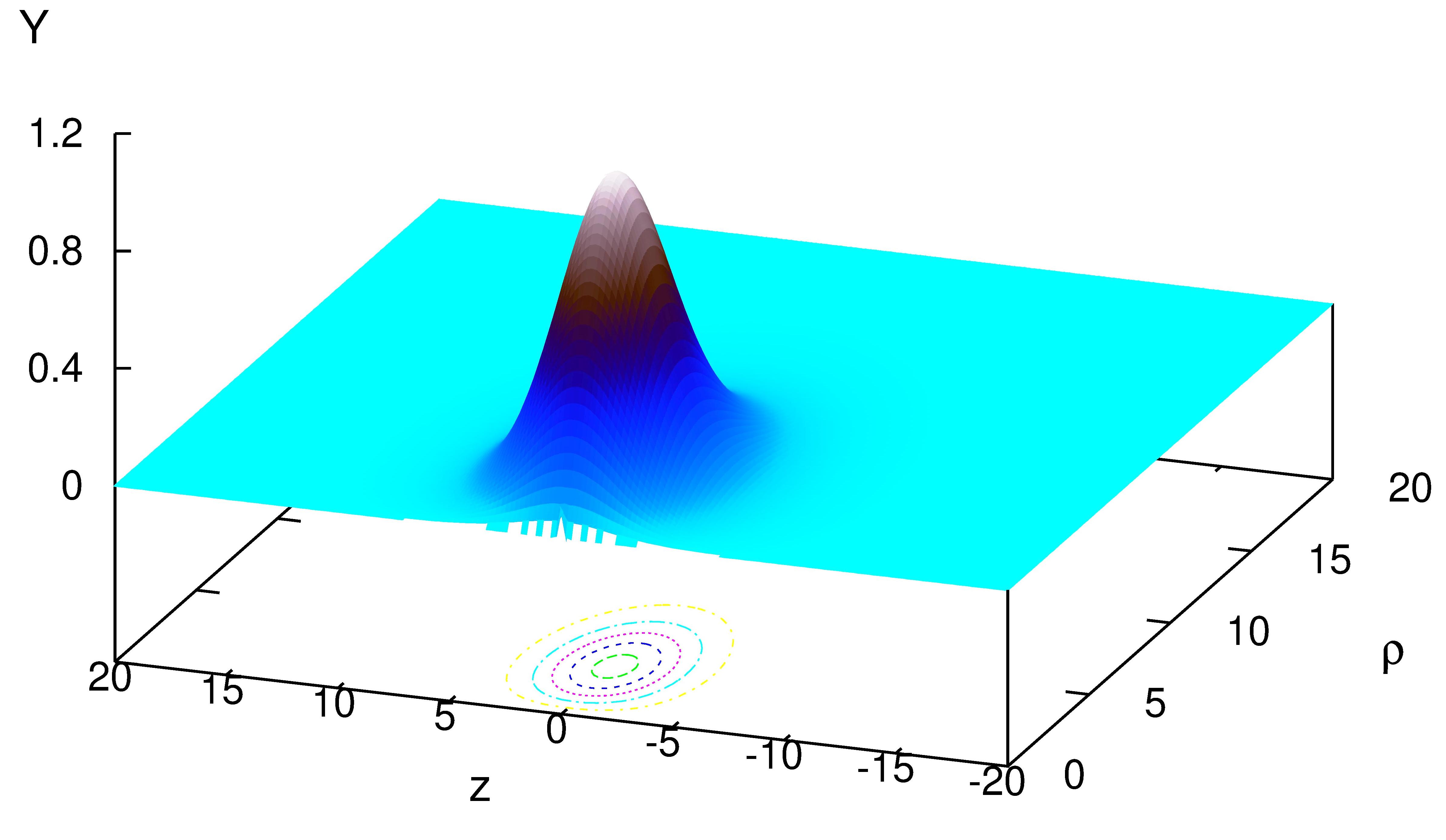}
\includegraphics[height=4cm]{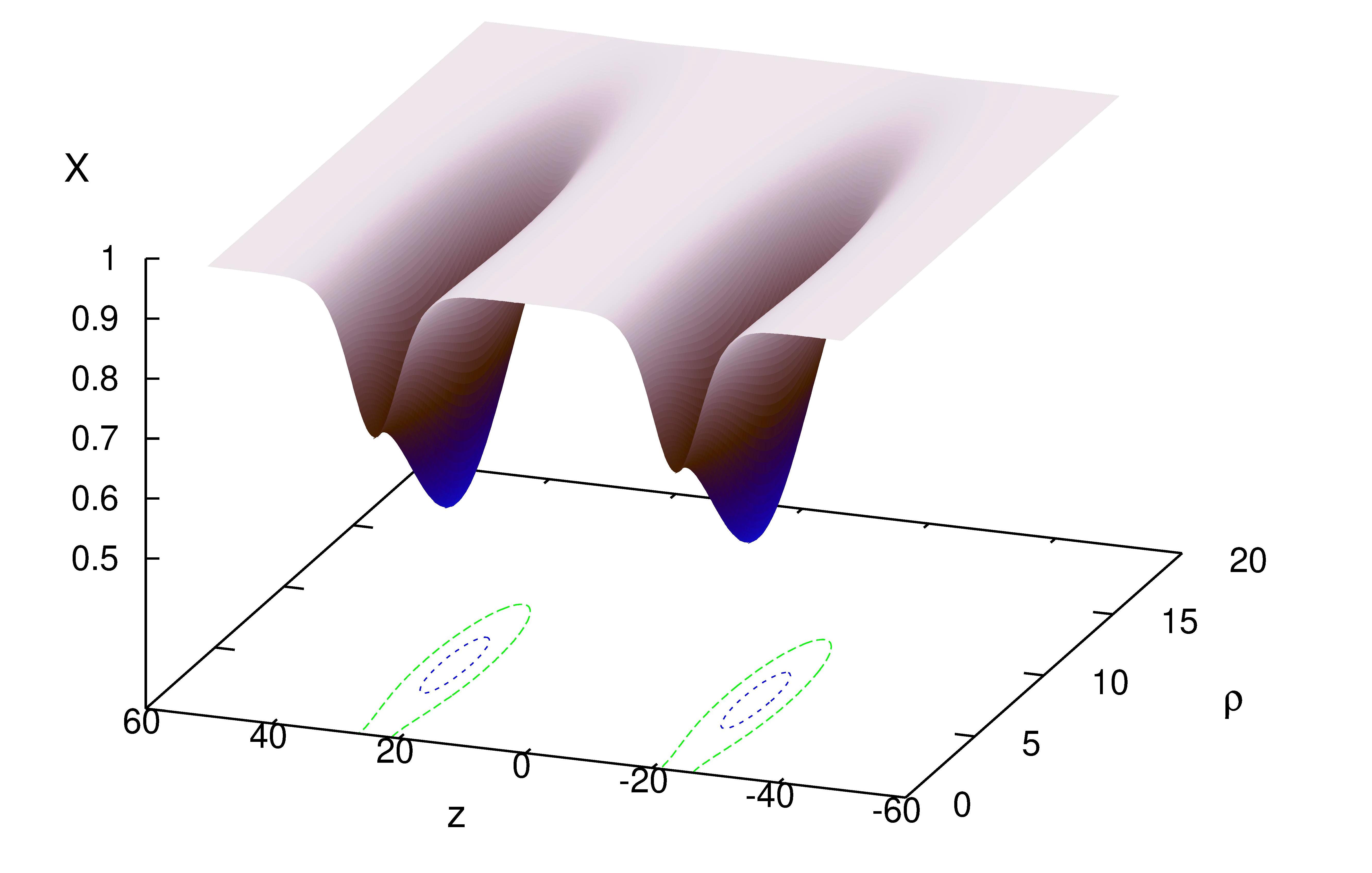}
\includegraphics[height=4cm]{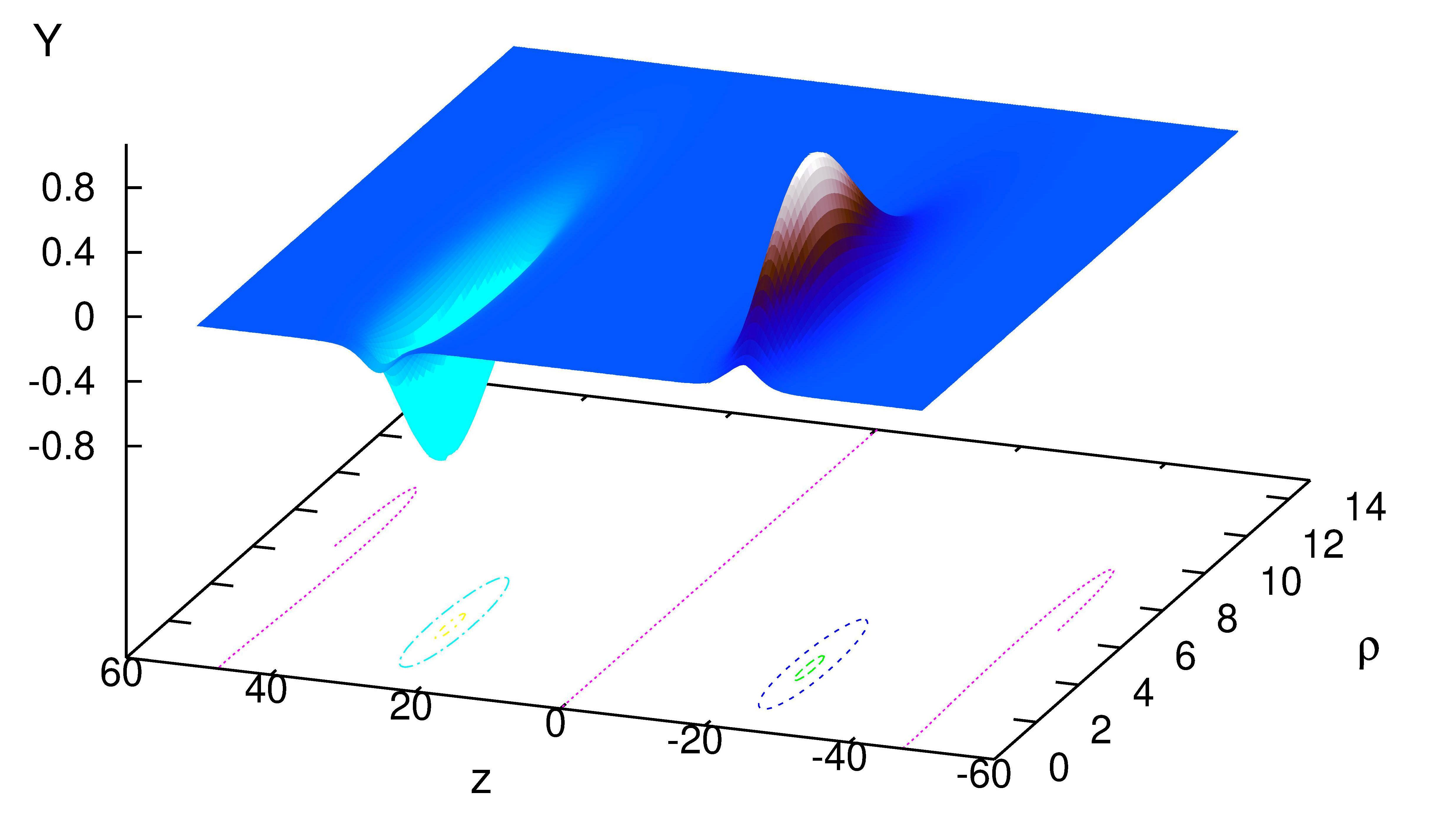}
\includegraphics[height=4cm]{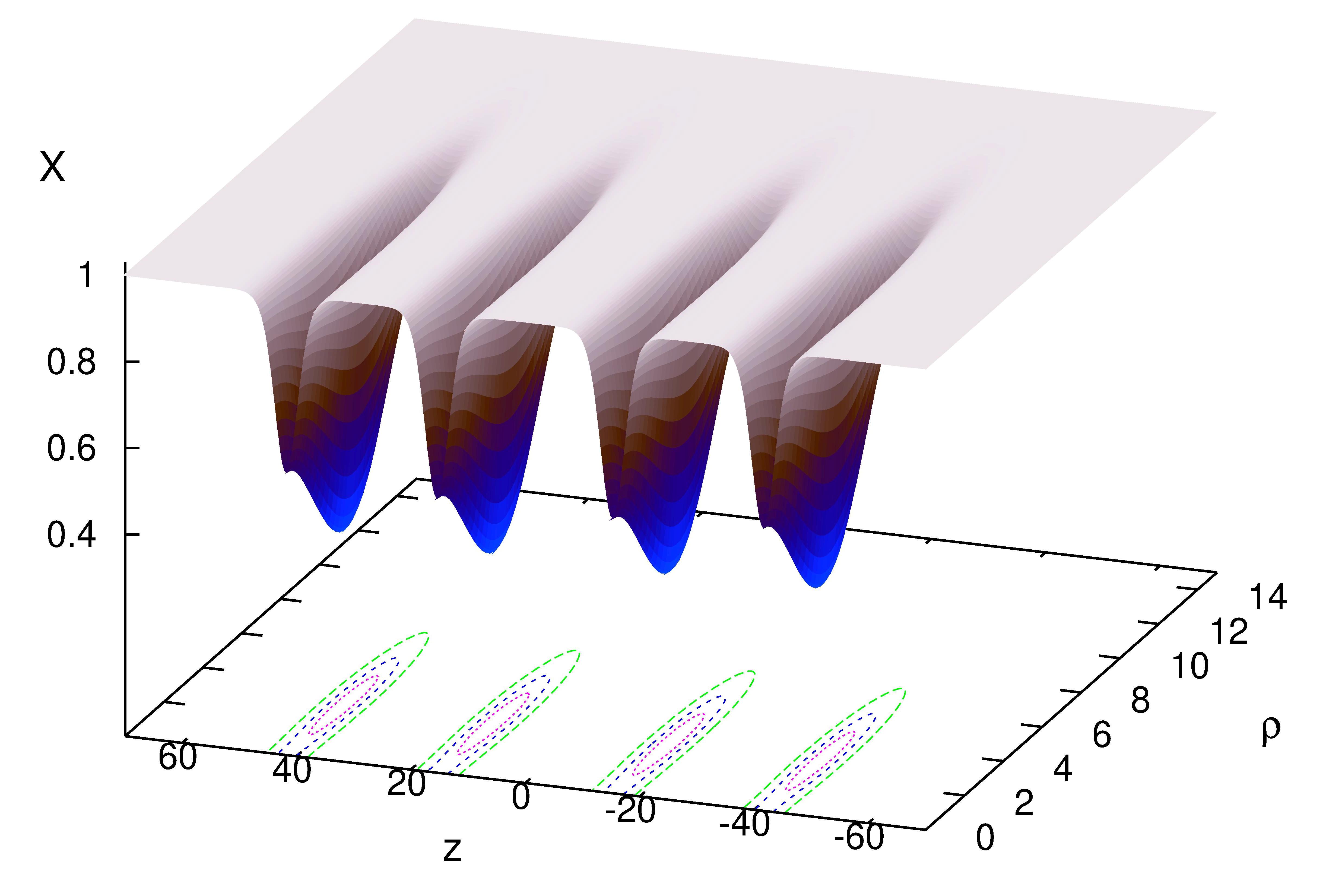}
\includegraphics[height=4cm]{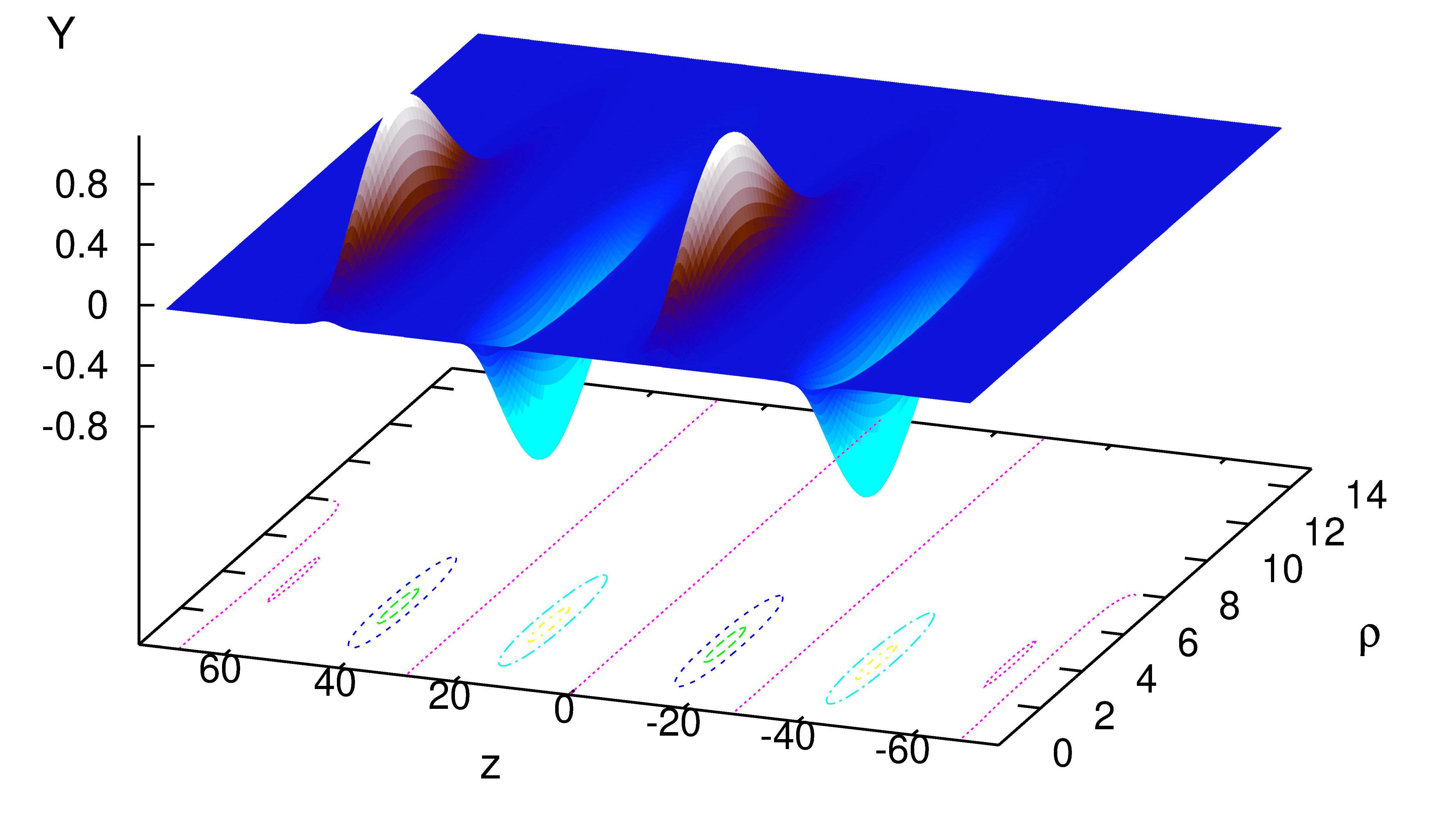}
\includegraphics[height=4cm]{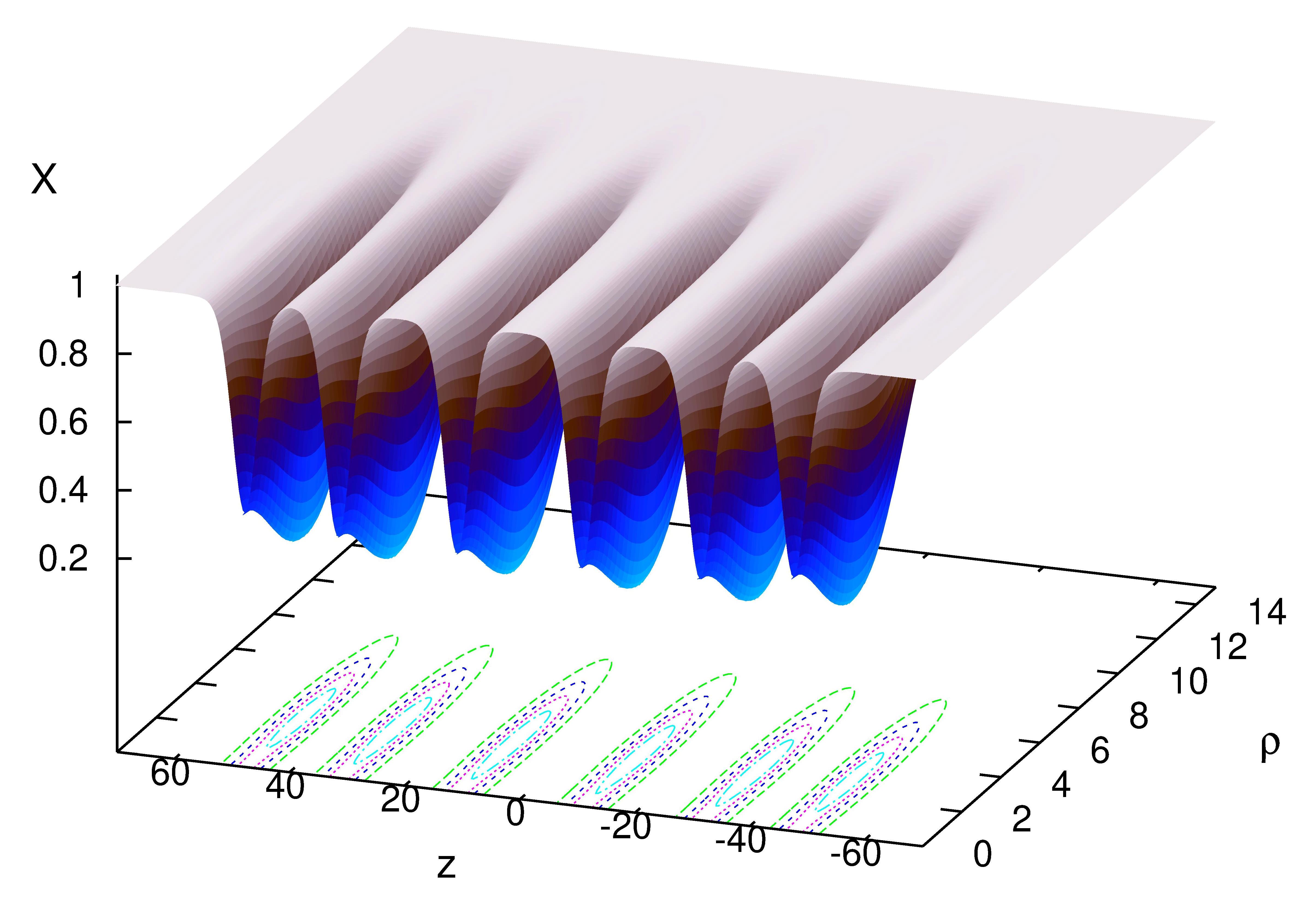}
\includegraphics[height=4cm]{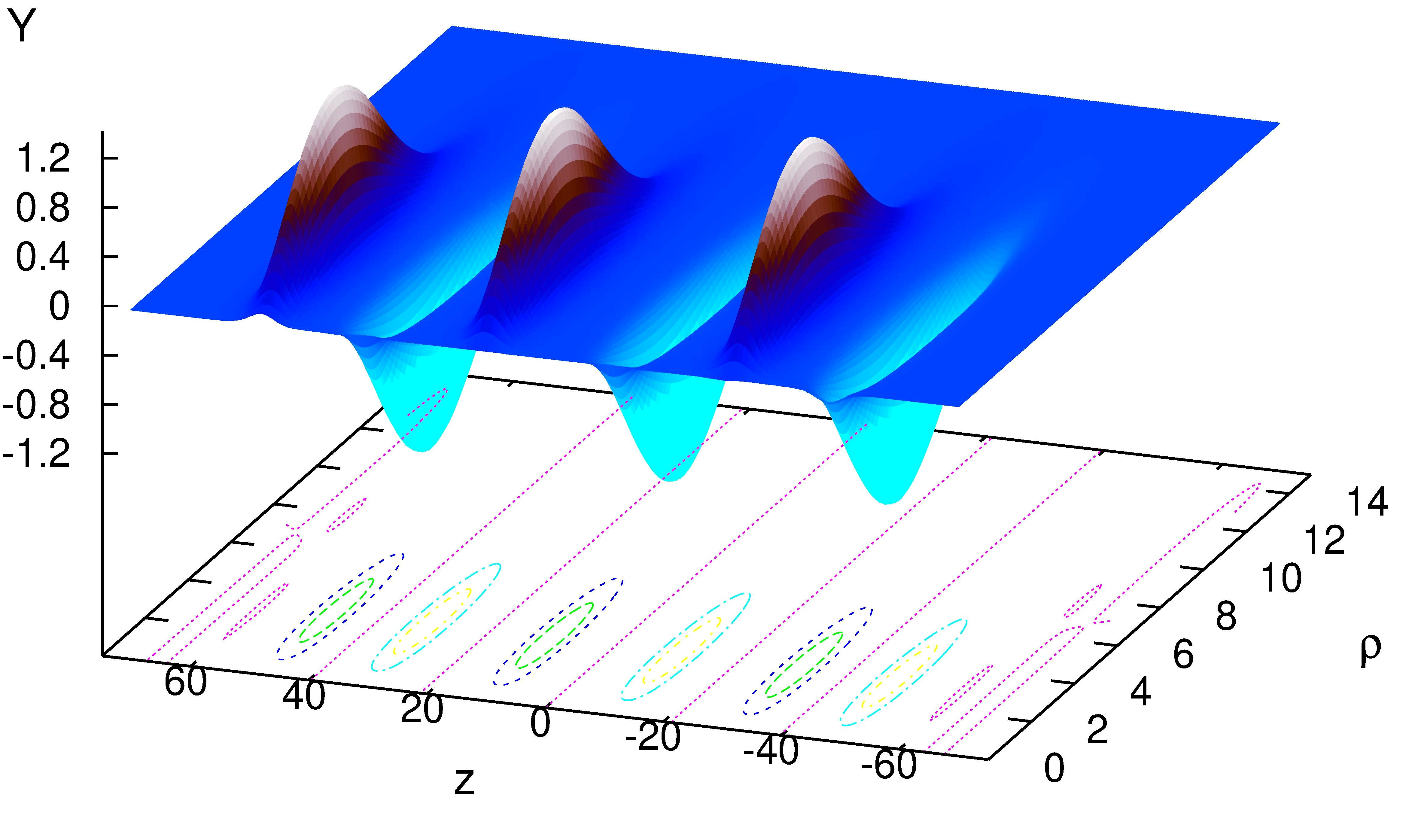}
\end{center}
\caption{\small
Axially-symmetric $n=1$ chains of gauged Q-balls:
The  field components  $X$ (left column)
and $Y$ (right column) of the $k=1,2,4$ and $k=6$
multisoliton configurations on the lower (electric) branch
are displayed at $\omega=0.90$ and $\mu=0.25$
and $g=0.07$.}
    \lbfig{fig1}
\end{figure}

\begin{figure}[h!]
\begin{center}
\includegraphics[height=3.1cm]{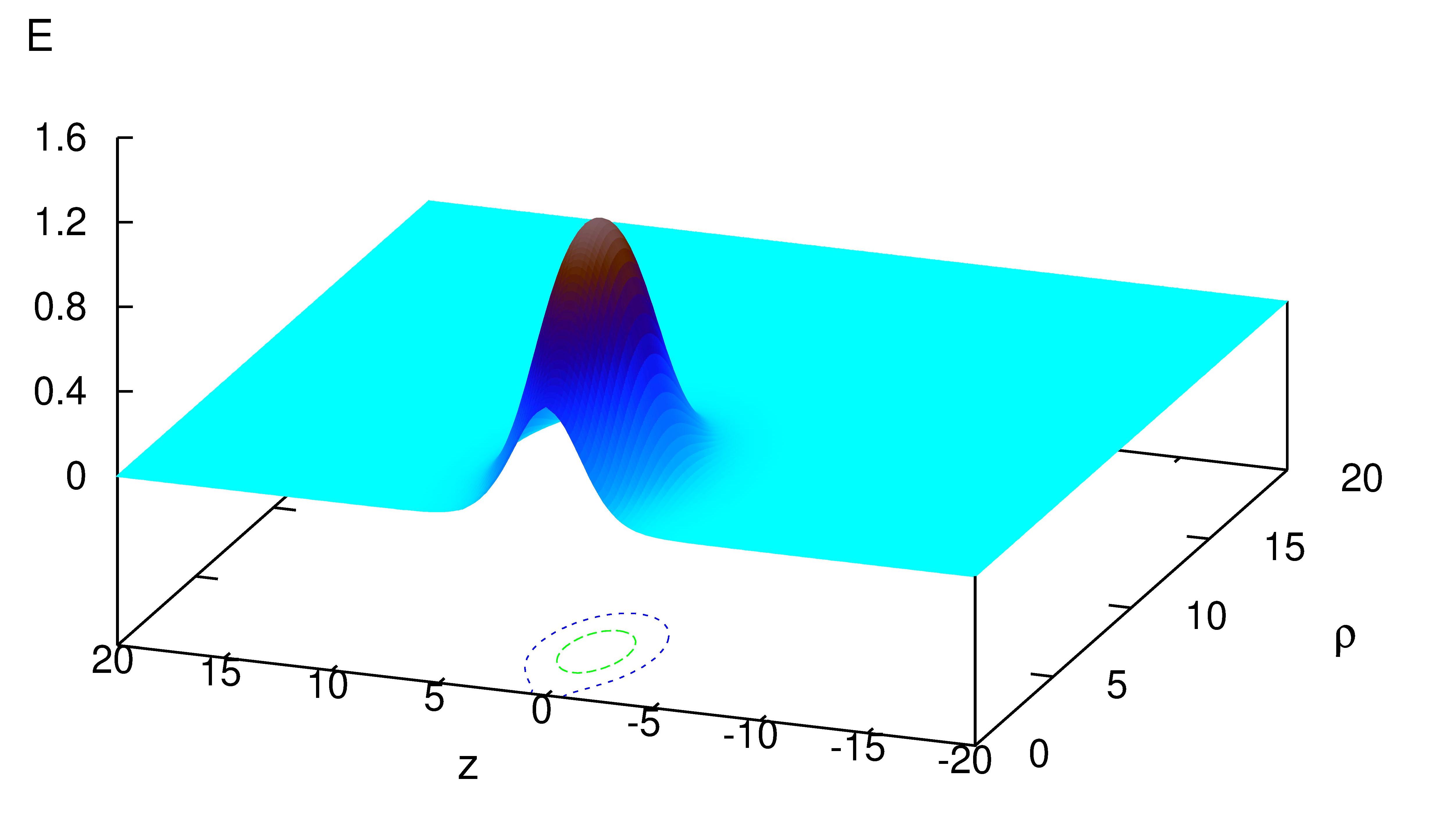}
\includegraphics[height=3.1cm]{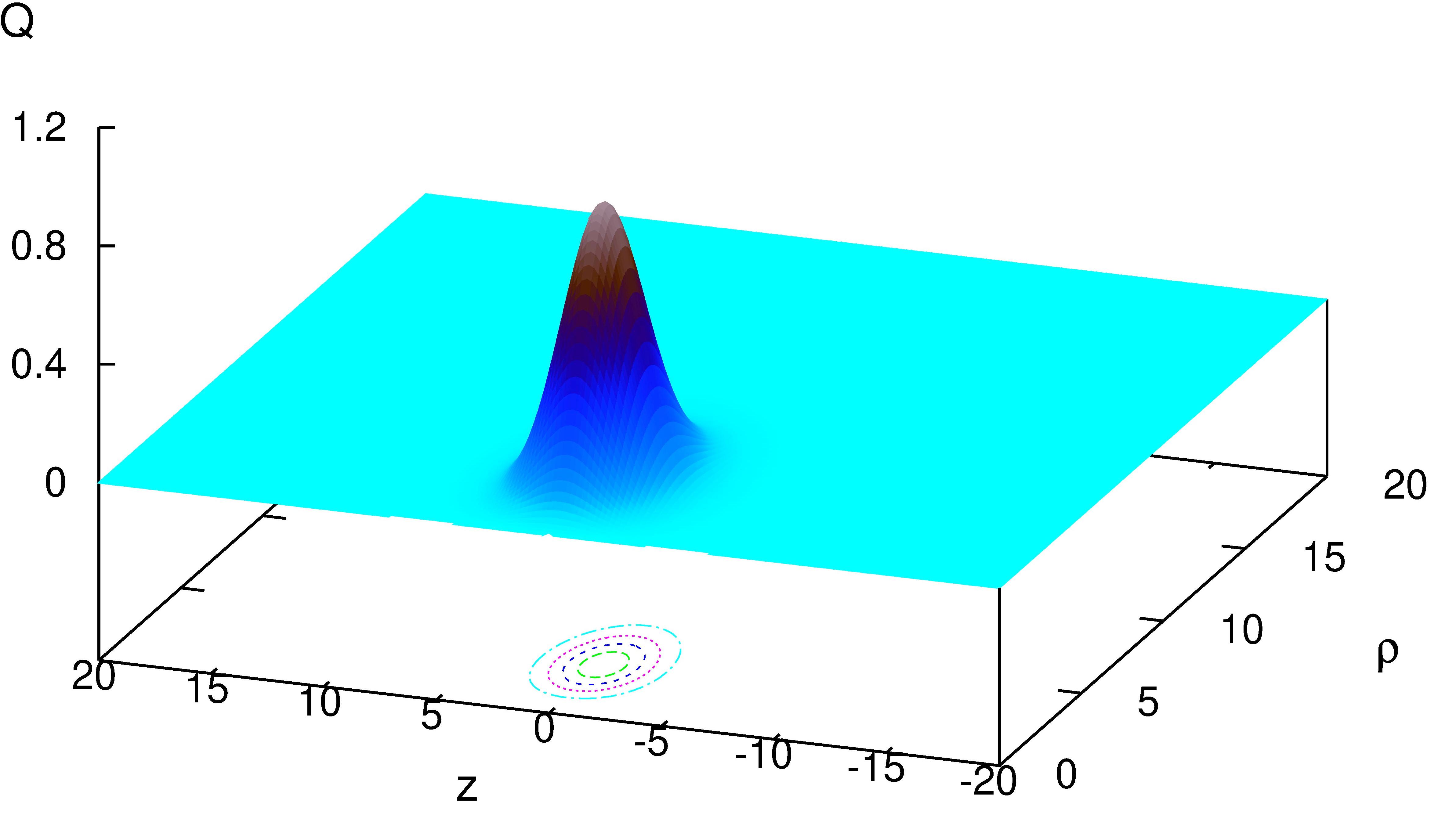}
\includegraphics[height=3.1cm]{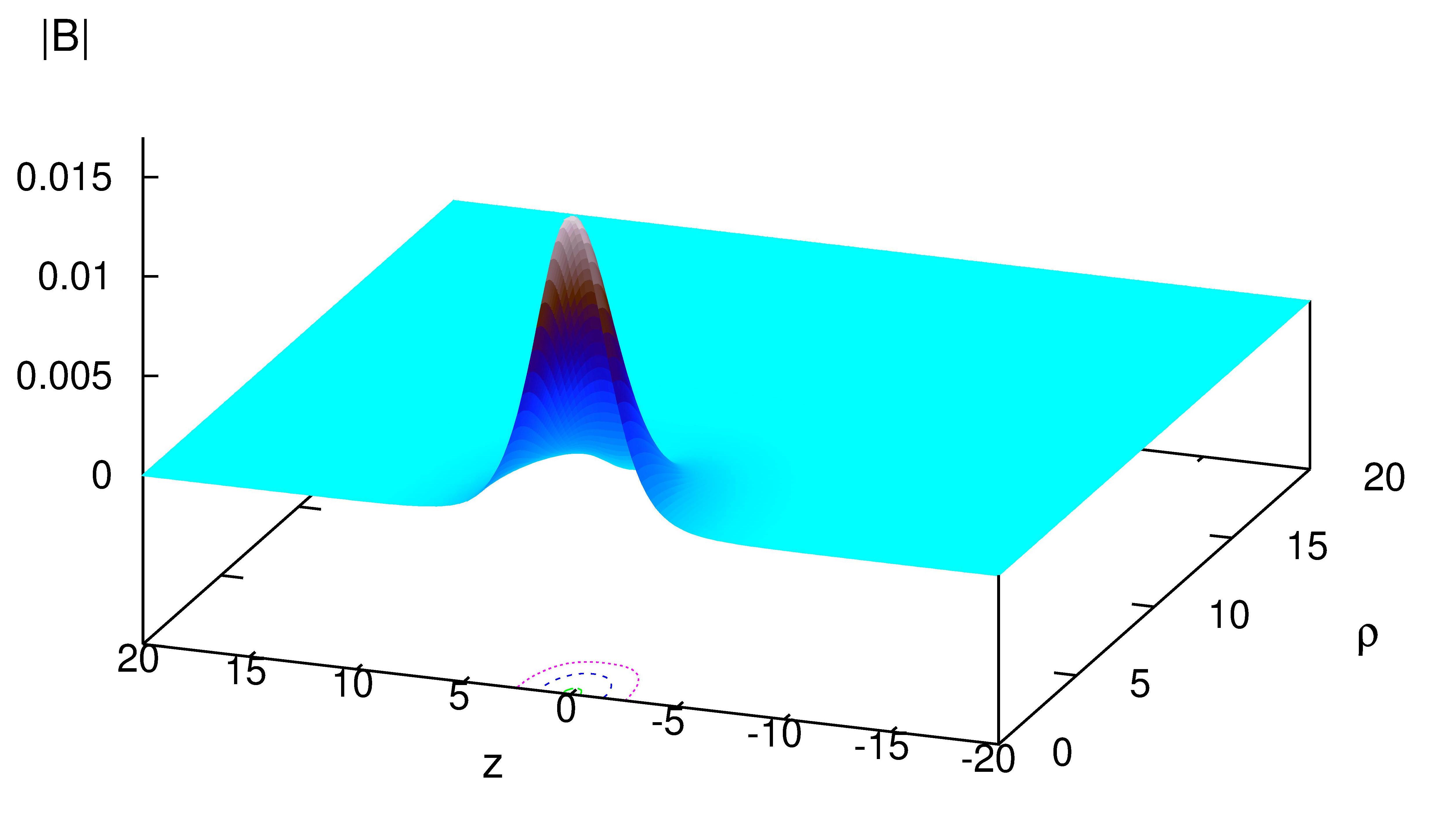}
\includegraphics[height=3.1cm]{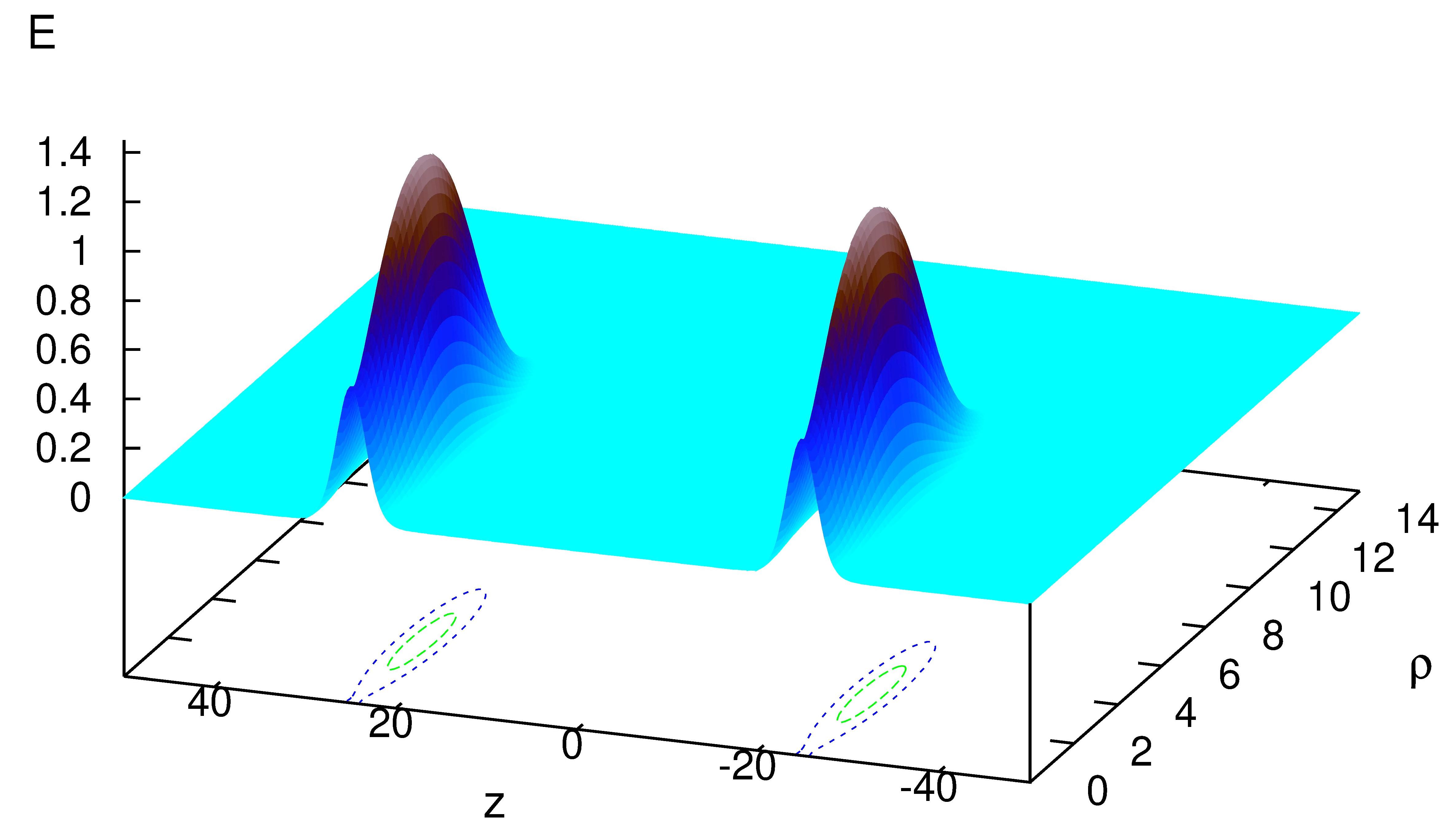}
\includegraphics[height=3.1cm]{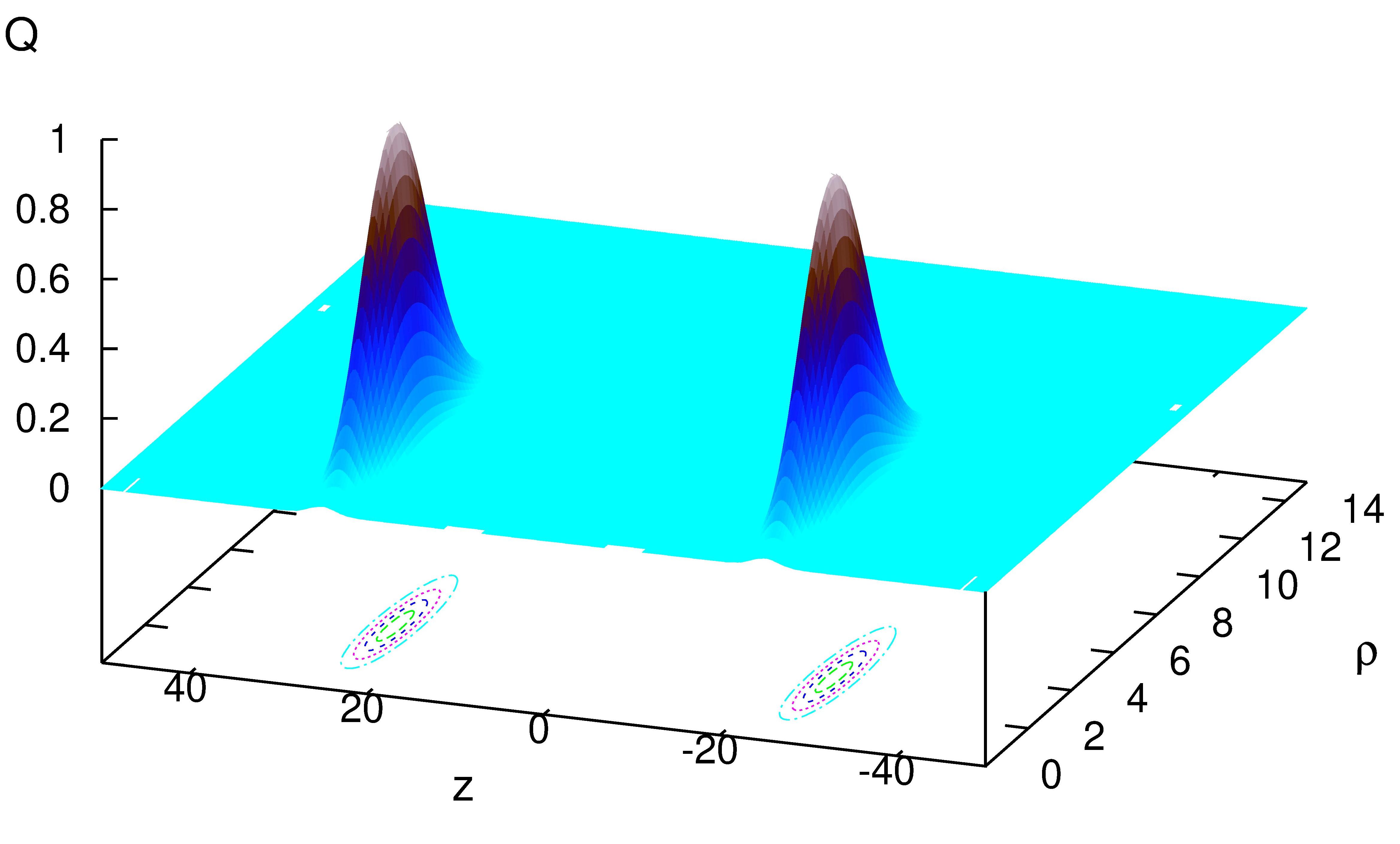}
\includegraphics[height=3.1cm]{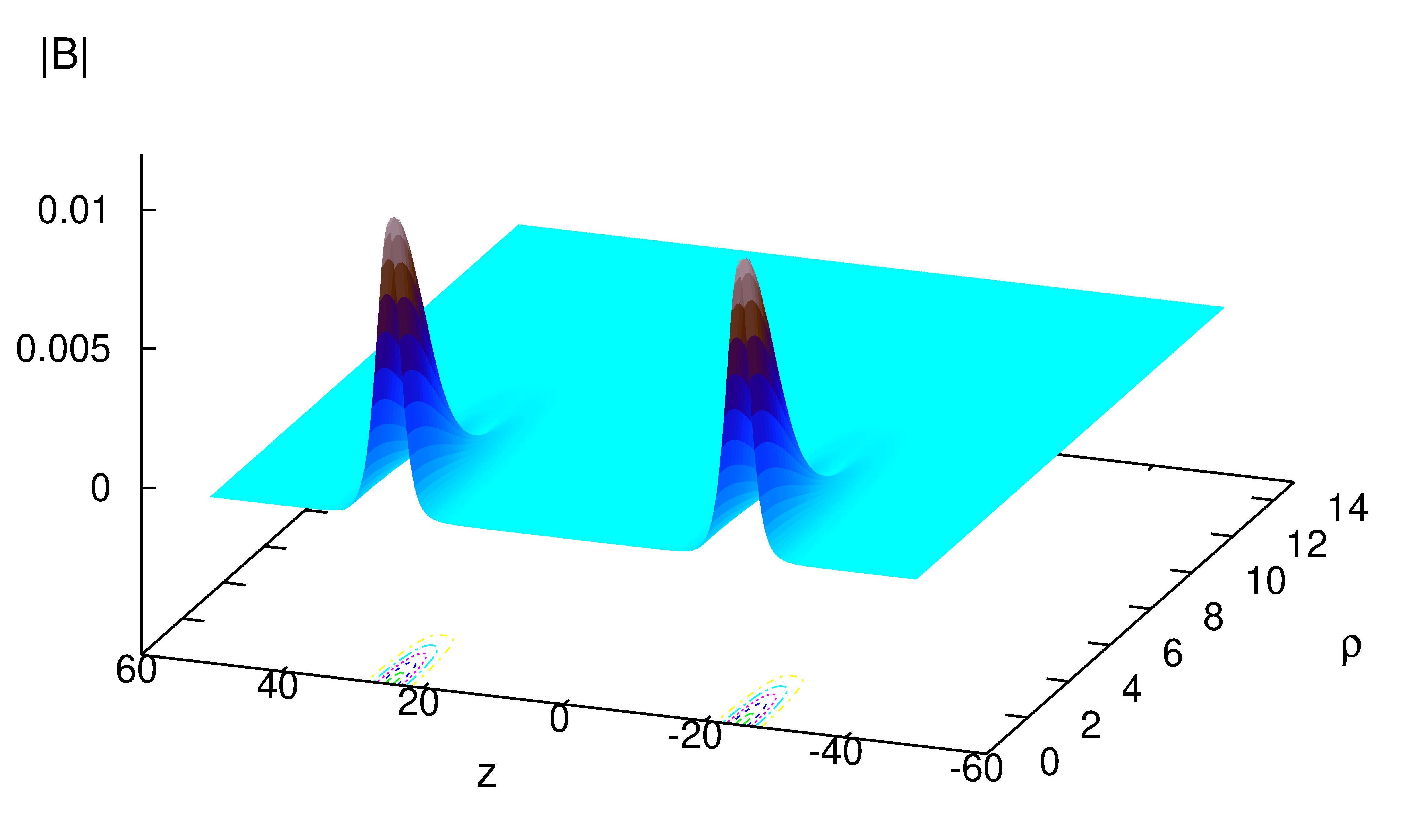}
\includegraphics[height=3.1cm]{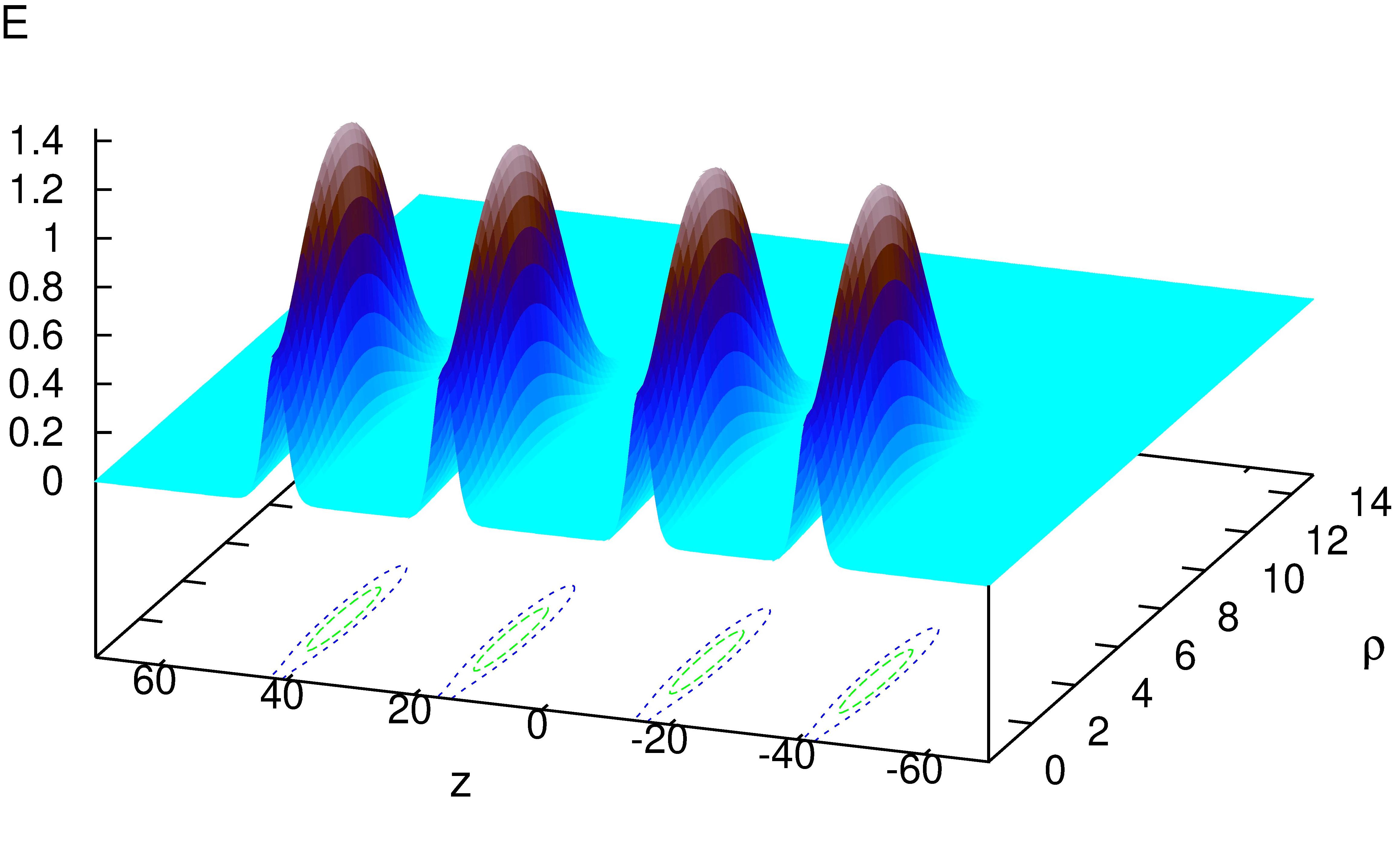}
\includegraphics[height=3.1cm]{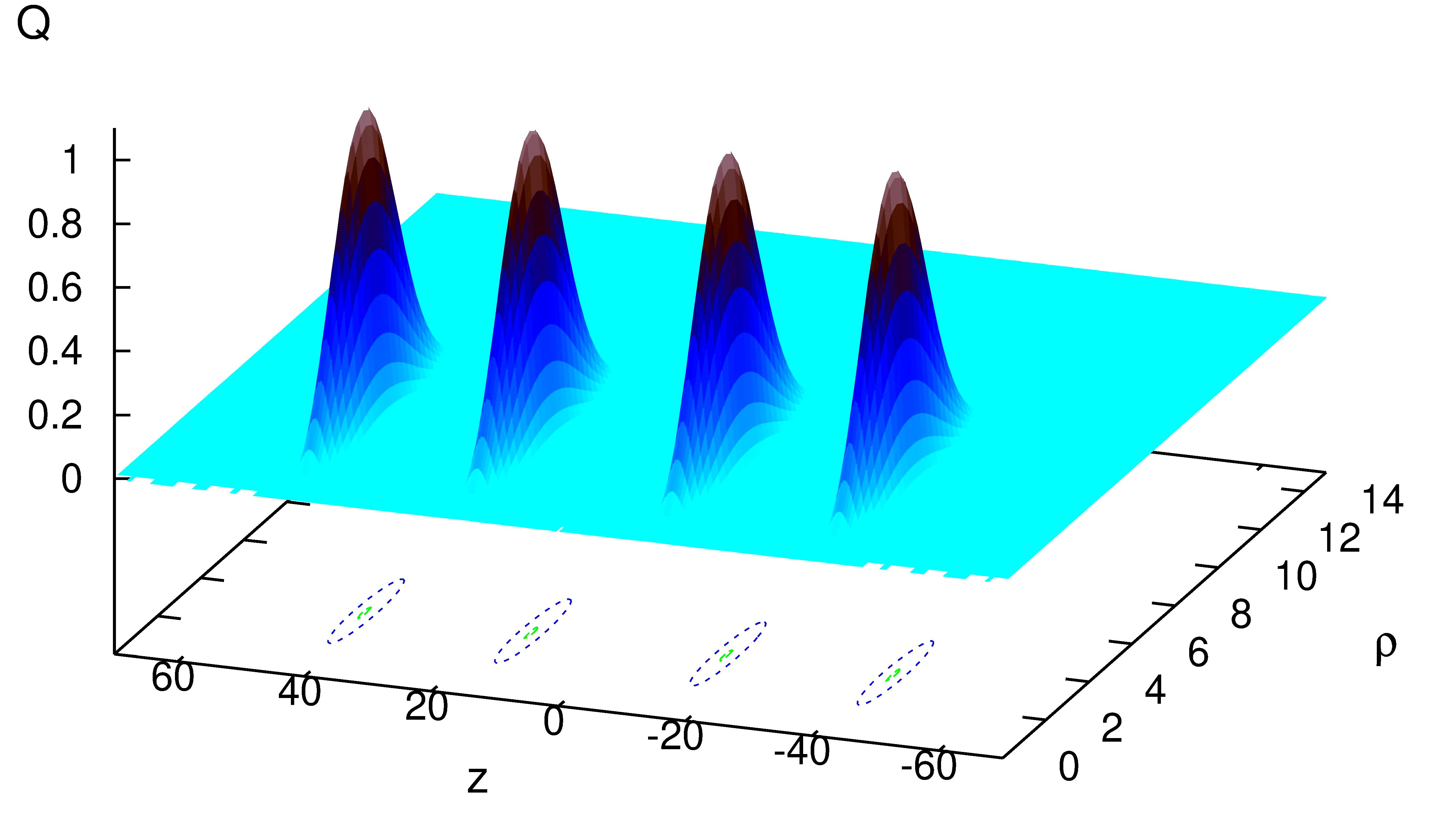}
\includegraphics[height=3.1cm]{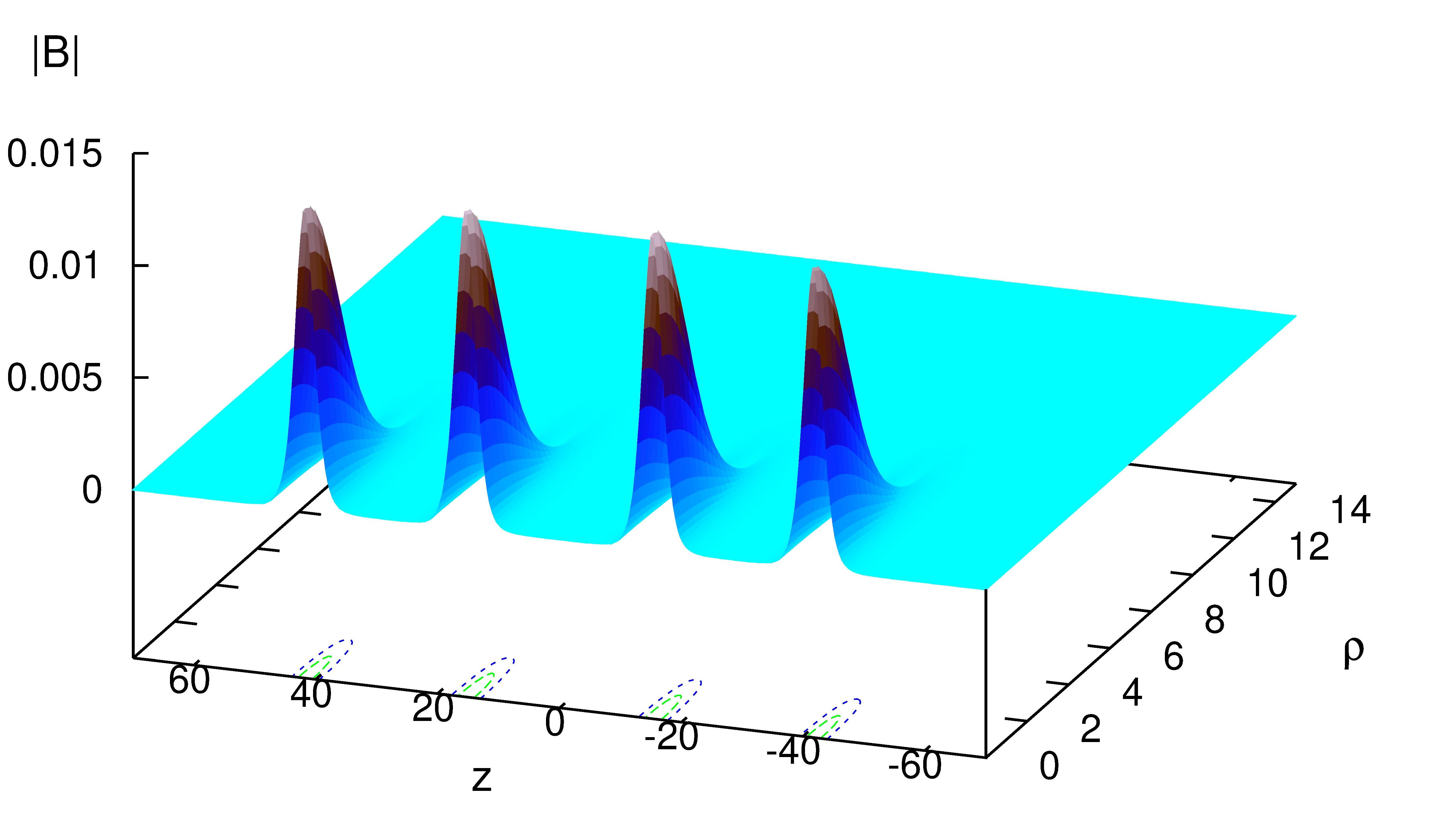}
\includegraphics[height=3.1cm]{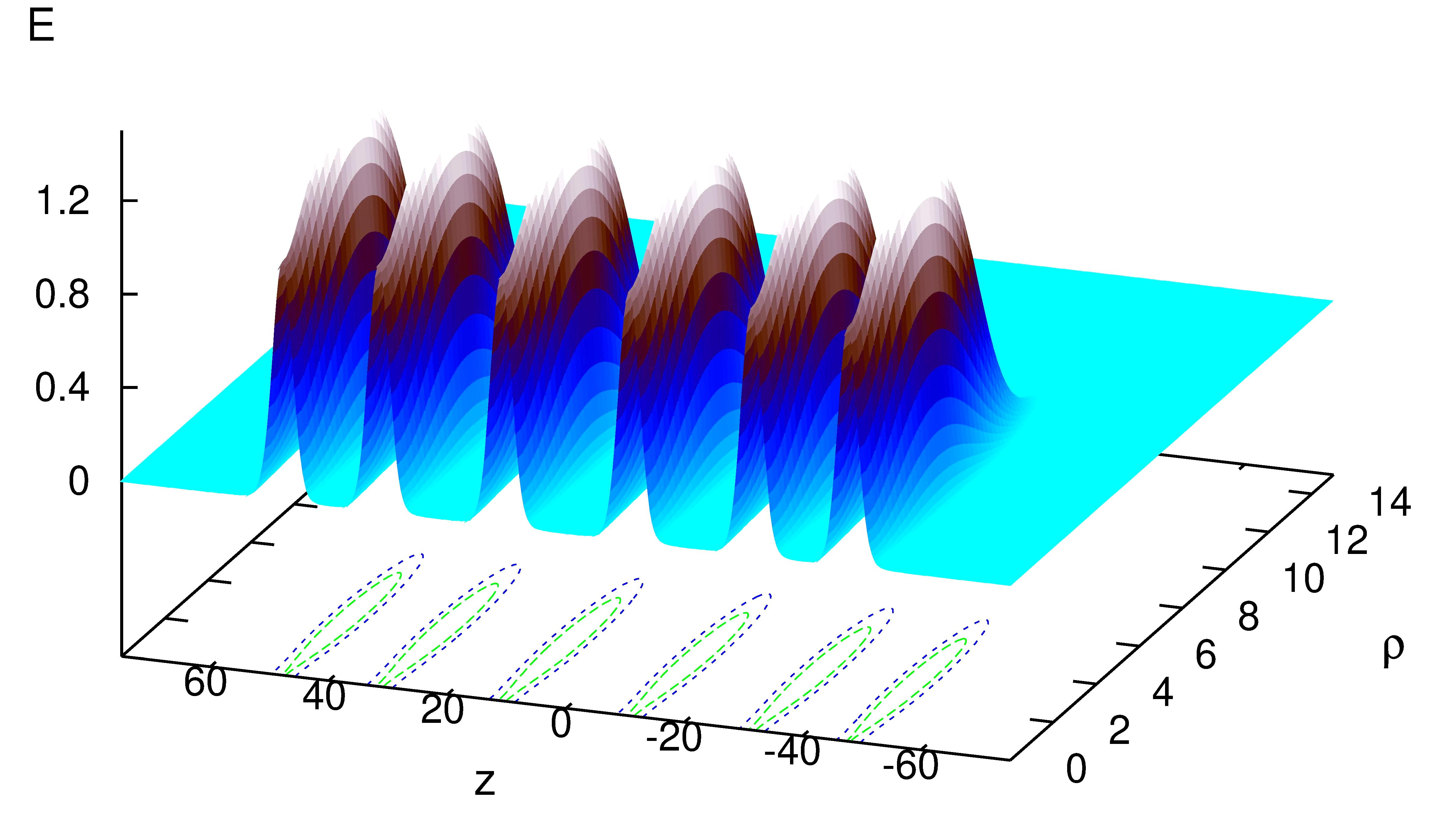}
\includegraphics[height=3.1cm]{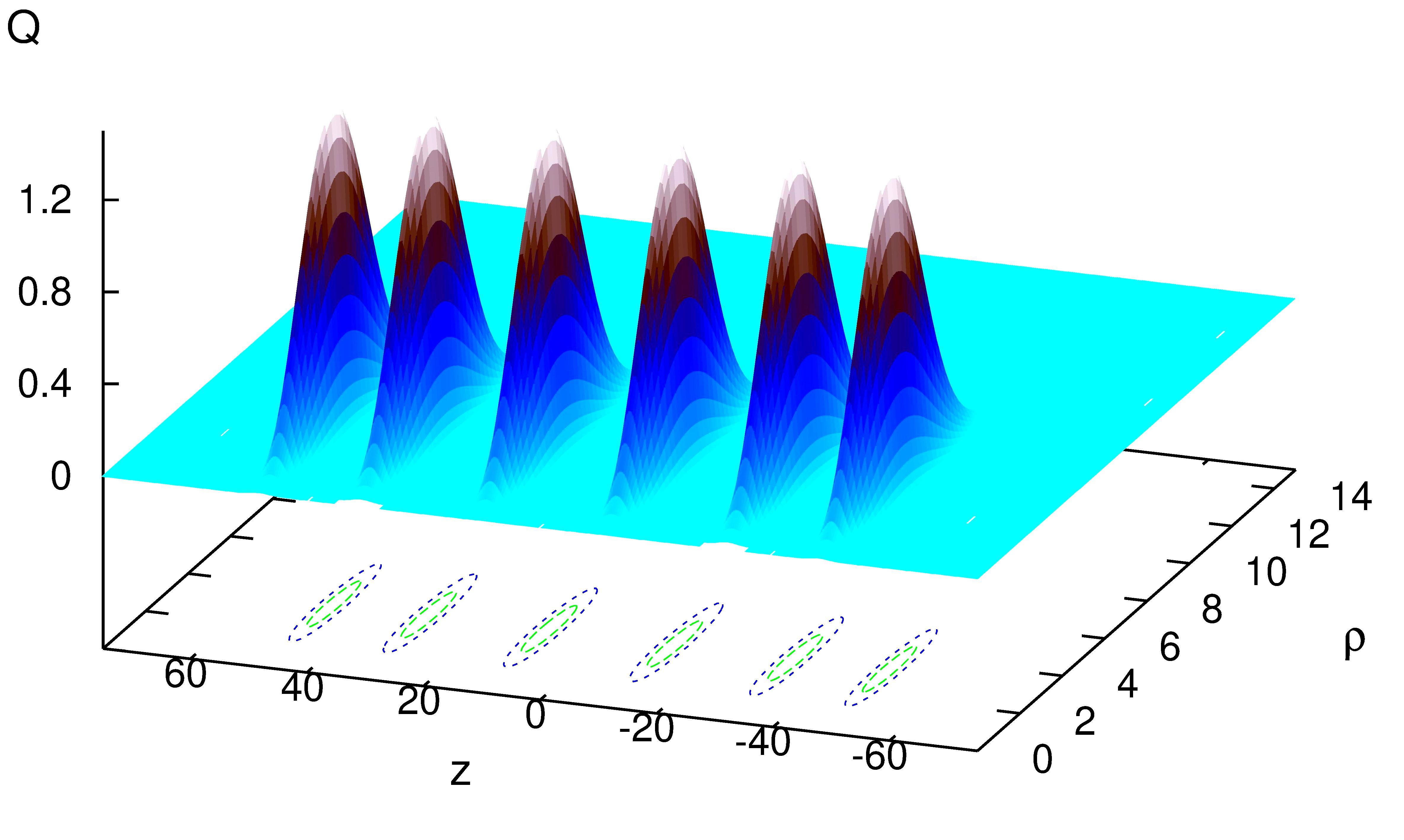}
\includegraphics[height=3.1cm]{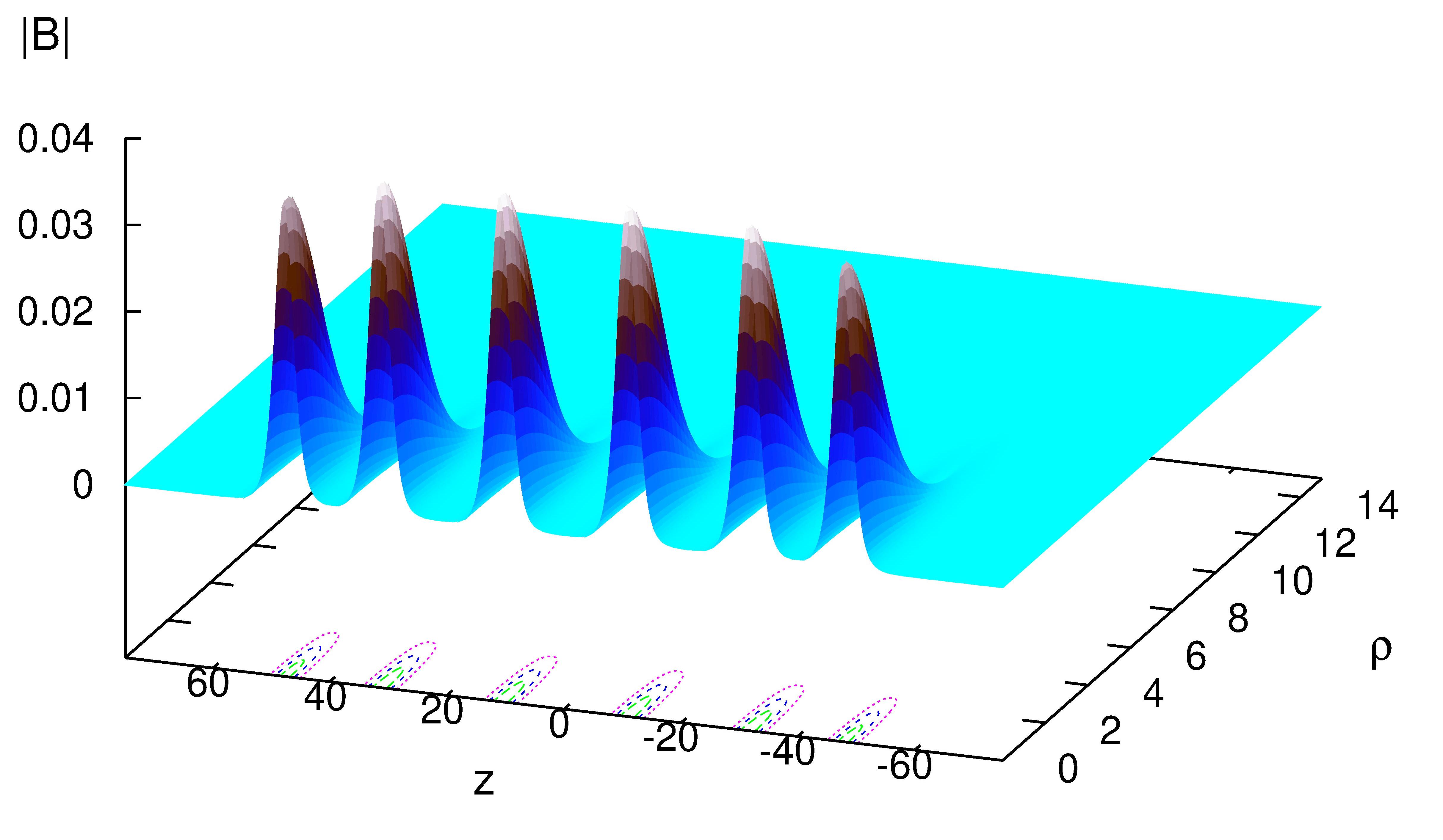}
\end{center}
\caption{\small
Axially-symmetric $n=1$ chains of gauged Q-balls:
The total energy density distribution   (left column),
the electric charge density distribution (middle column),
and the magnitude of the magnetic field distribution (right column)
of the $k=1,2,4$ and $k=6$ multisoliton configurations on the lower (electric) branch
are displayed at $\omega=0.90$ and $\mu=0.25$
and $g=0.07$.}
    \lbfig{fig2}
\end{figure}
Simplest spherically symmetric solutions of the gauged FLS model
were constructed a long time ago \cite{Lee:1991bn}. These  $n=0$  gauged Q-balls may
exist in some restricted domain of values of the
parameters of the system. The repulsive electrostatic Coulomb interaction reduces the allowed range
of values of the angular frequency of the spinning gauged Q-ball, the minimal value $\omega_{min}$ is increasing
as the gauge coupling $g$ increases. As in the decoupled limit,
the values of the scaled frequency $\omega$ are also bounded from above by the mass of the complex scalar field,
$\omega\le 1$. However, the $U(1)$ gauged Q-balls may exist as a localized finite mass
field configuration even in the limit $\omega=1$ \cite{Gulamov:2015fya}. Further, radially excited $U(1)$
gauged Q-balls we studied recently in \cite{Loginov:2020xoj}, the angularly excited parity-even
solutions of the Friedberg-Lee-Sirlin-Maxwell model were considered in out previous work \cite{Loiko:2019gwk}

Here we study new type of axially symmetric parity-odd solutions of the model \re{lag-fls}, which can be considered as
composite configurations, the chains of spinning gauged Q-balls, located
symmetrically with respect to the origin along the symmetry axis. These solutions can be classified by the
winding number $n$ and
the number of constituents $k$. A simple non-trivial configuration of that type  represents
a $k=2$ pair of $n=1$ Q-balls spinning in opposite phases, as shown in Fig.~\ref{fig1}, second row.
The pair is stabilized by the magnetic field generated by the circular
current. Our numerical calculations show that there is no similar
axially-symmetric $n=0$  solution \footnote{Notably, the chains of
solitons with zero angular momentum, stationary spinning on the symmetry axis, may exist in curved spacetime
\cite{Herdeiro:2020kvf}.}.

A few chain solutions of that type on their fundamental branch for a given value of the
mass parameter $\mu=0.25$ and angular
frequency $\omega=0.90$ are exhibited in Figs.~\ref{fig1},\ref{fig2}.
The plots in the Fig.~\ref{fig1} represent
the scalar field functions $X(r,\theta), Y(r,\theta)$ of the $k=1,2,4,6$ chains.
The first row of Fig.~\ref{fig1} shows the field functions of a single $n=1$ parity-even gauged Q-ball for comparison.
For clarity, we have chosen polar coordinates
$\rho = r\sin \theta$ and $z=r\cos\theta$ in the figures.
These Q-chains possess $k$ constituents,
as nicely seen by the number of peaks of the distributions of the total energy density,
the charge density and the magnitude of the
magnetic field, as seen in the Fig.~\ref{fig2}. Note, that, apart parity-even $k=1$ circular vortex,  we do not  obtain
chains with odd number of constituents,  although we cannot exclude a possibility these solutions may exist for
higher values of the asimuthal winding number $n$.

Both for parity-even and for parity-odd  axially symmetric  solutions of the gauged model \re{lag-fls}
the angular frequency $\omega$ is bounded from above. The gauged spinning solutions exist within a frequency interval
$\omega_\mathrm{min}\le \omega \le \omega_\mathrm{max}$, the lower critical value of the frequency $\omega_\mathrm{min}$
depends on  the electromagnetic coupling.

The spinning gauged Q-balls arise as perturbative excitations as
the angular frequency is decreasing slightly below the mass threshold  $\omega_\mathrm{max}=1$, see
Fig.~\ref{fig3}. The $k$-chains of spinning Q-balls form the first branch of solutions, the angular frequency is decreasing
along this branch. The constituents with non-zero angular momentum posses both the electric charge and
circular magnetic field, which is generated by the Noether current $j_\mu$ \re{Noether}.
The corresponding solenoidal magnetic field has a set of $k$ pronounced maxima equidistantly located
on the symmetry axis, while the electric charge of the constituents is pushed outwards, as seen in Fig.~\ref{fig2}.

\begin{figure}[t]
\begin{center}
\includegraphics[width=.32\textwidth, angle=-90]{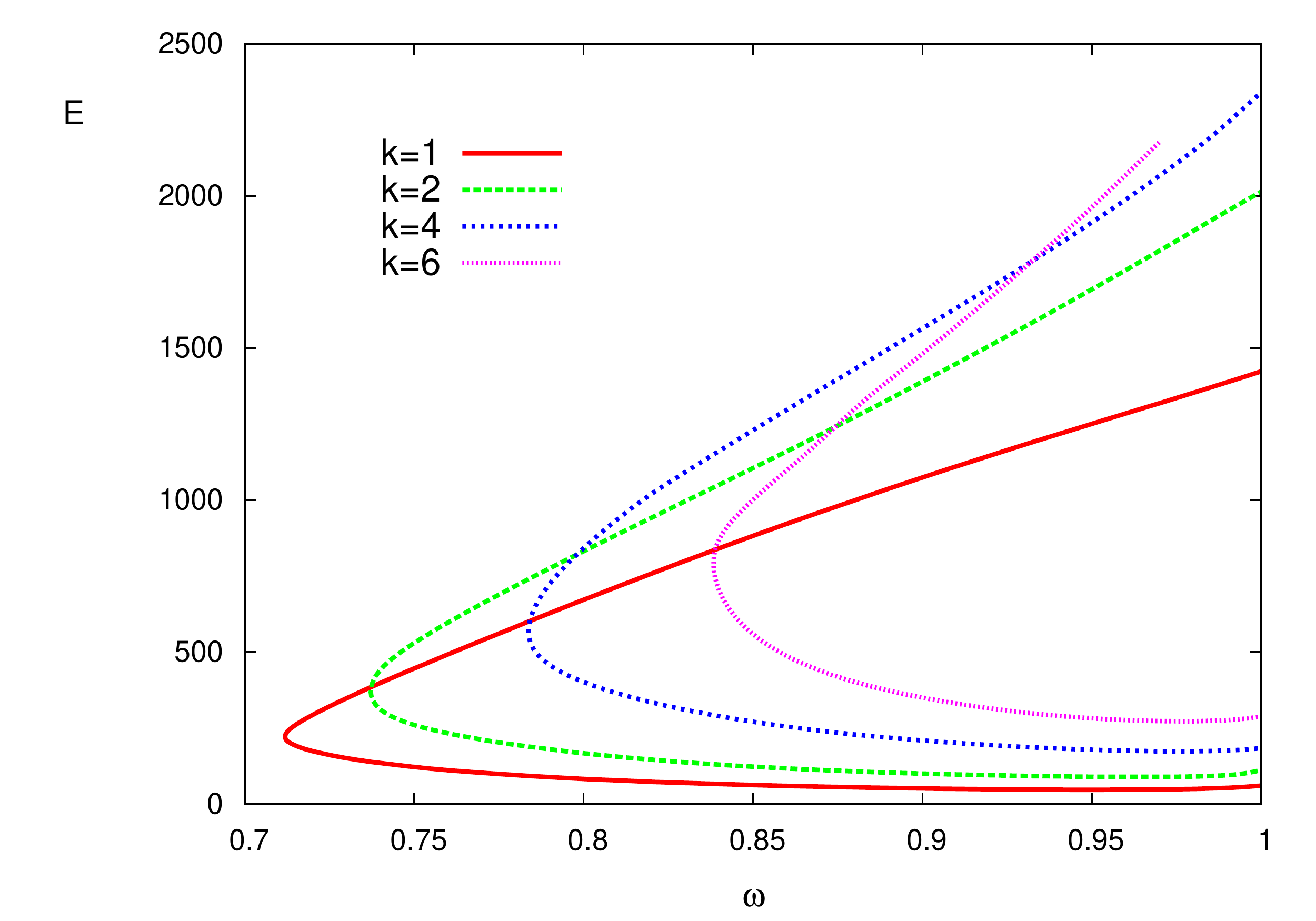}
\includegraphics[width=.32\textwidth, angle=-90]{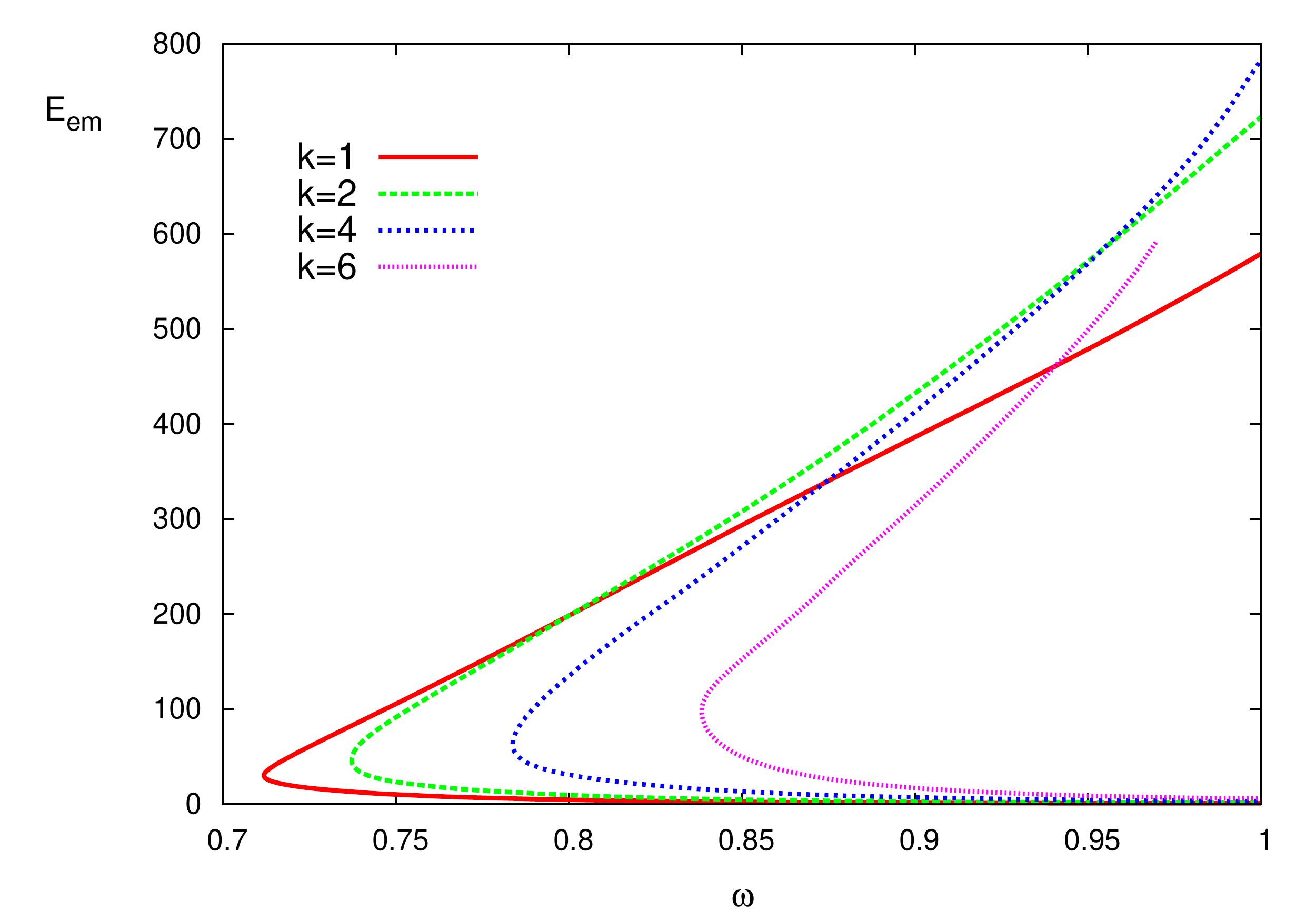}
\end{center}
\caption{\small
Axially-symmetric $n=1$ chains of gauged Q-balls:
The total energy $E$ (left plot) and electromagnetic energy $E_{em}$ (right plot)
are shown as functions of the angular frequency $\omega$ for the  parity-even $(k=1)$ solution,
the parity-odd pair of Q-balls ($k=2$) and for the $k=4$ and $k=6$ chains,
at $g=0.07$ and $\mu=0.25$}.
\lbfig{fig3}
\end{figure}

\begin{figure}[t]
\begin{center}
\includegraphics[width=.32\textwidth, angle=-90]{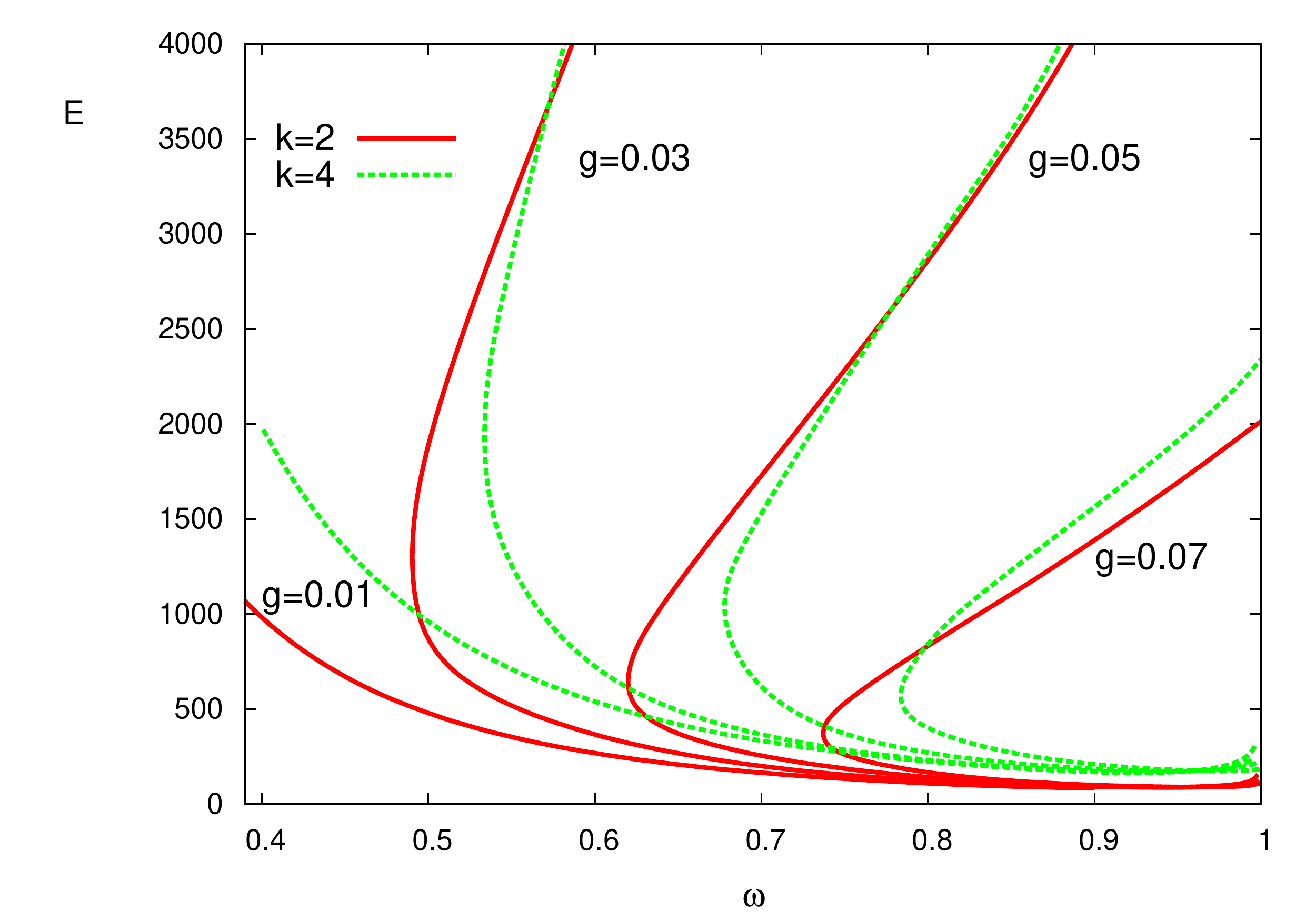}
\includegraphics[width=.32\textwidth, angle=-90]{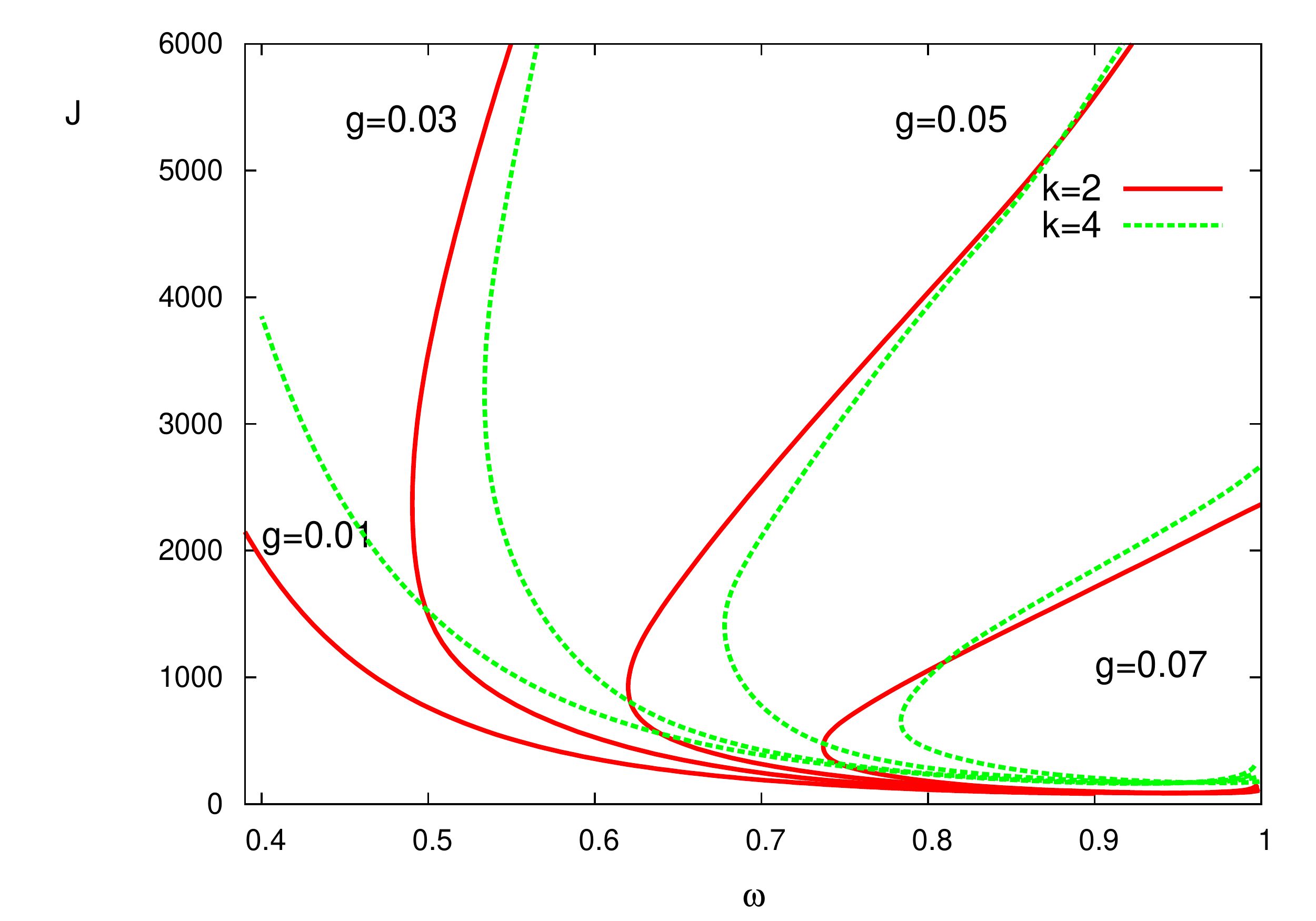}
\end{center}
\caption{\small
Axially-symmetric $n=1$ chains of gauged Q-balls:
The total energy $E$ (left plot) and the angular momentum $J$ (right plot) of the parity-odd $k=2$ pair
and 4-chain of the Q-balls are displayed as functions
of the angular frequency $\omega$
for some set of values of the gauge coupling $g$ at $\mu=0.25$.}
\lbfig{fig4}
\end{figure}

The electromagnetic energy of the spinning solitons remains relatively small on the fundamental branch, there
the electric Coulomb force is balanced mainly by the scalar interaction while the magnetic field is comparatively weak.
As for the parity-even gauged Q-balls, we refer to that branch to as "electric" one \cite{Loiko:2019gwk}.
Since the local $U(1)$ symmetry becomes broken in the interior of the constituents of a chain, the
electric branch corresponds to the "superconductive" phase.

Considering the frequency dependence of $n=1$ $k$-chains of gauged Q-balls, we found that it is qualitatively the same as
for the parity-even spinning solutions, we discussed in \cite{Loiko:2019gwk}.
Both the size and the electric charge of the constituents of the chain increases
as the angular frequency is decreasing from the mass threshold  $\omega_\mathrm{max}=1$.
Hence, both the current $j_\mu$ and associated magnetic field $B$,
become stronger. For some critical value of the frequency
$\omega_\mathrm{min}$ the value of the real component of the configuration
approaches some set of zeros,
then the electromagnetic field becomes massless on some set of $k$ circular domains in space around the symmetry axis.
As a result, the symmetry is restored
in these regions and the
energy of the magnetic field  becomes higher than the electrostatic energy of the spinning Q-ball. In other words, the second,
magnetic branch is formed \cite{Loiko:2019gwk}. The energy of the configuration, as well as the change $Q$ and
the angular momentum $J$  are rapidly increasing with the
angular frequency, the magnetic branch extends forward, as shown in Figs~\ref{fig3},\ref{fig4}.

Further increase of the frequency leads to expansion of the domains of normal phase in a chain,
where the real component becomes trivial, $\psi=0$,
and both the electromagnetic and complex scalar field are massless.
The critical value of the frequency $\omega_\mathrm{min}$ depends on the gauge coupling $g$, see Figs~\ref{fig4}, it
increases with $g$, as the electric and magnetic branches become shorter.
It also increases
for larger chains with higher number of constituents $k$ as the gauge coupling $g$ remains fixed, see Figs~\ref{fig3}
where we exhibit the total energy $E$ and the energy of the electromagnetic field $E_{em}$
of the gauged $n=1$ k-chains as functions of the angular frequency $\omega$ at $g=0.07$. The separation between the constituents
of the chain on the magnetic branch decreases as $\omega$ increases, the chain becomes energetically
unstable, see Fig.~\ref{fig3}.

Note that, both the for parity even and parity-odd solutions,
the magnetic branch may exist only for a non-zero values of the mass parameter $\mu$, it extends forward almost linearly
with $\omega$. When $\mu$ becomes higher, the critical value of the frequency $w_{cr}$ is increasing \cite{Loiko:2019gwk}.

Finally, we note that, similar to the case of the usual Q-balls, both the energy and the charge of the
gauged $k$-chains are minimal at some critical value of the frequency
$\omega_{cr} > \omega_\mathrm{min}$. This is an indication of a
cusp in the curve of $E(Q)$ dependency\footnote{Note that for the gauged spinning Q-balls a
multi-cusp pattern is generally observed \cite{Loiko:2019gwk,Loginov:2020lwg}.}, i.e., the existence of
different solutions with the same value of charge $Q$. Hence, one can expect the more energetic
gauged $k$-chains on the upper magnetic branch are unstable.
More general, the multi-tori  structure of the $k$-chain solutions, which represent saddle points of the
energy functional, is energetically unfavorable
as compared to the single torus-like structure of the parity-even solutions.

\section{Conclusions}

The main purpose of this work was to present a new type of
axially-symmetric solutions of the $U(1)$ gauged FLS model,
which represent electromagnetically
bounded chains of even number of stationary spinning charged Q-balls with non-zero angular momentum coupled
with a solenoidal magnetic field.

The chains emerge from the perturbative fluctuations at a maximal value of the angular frequency,
given by the mass of the real scalar field. They then ascend along their fundamental electric branch,
until a minimal value of the frequency is reached, which, for a fixed value of the mass parameter, is determined by
the gauge coupling strength $g$. Then the second, magnetic branch is formed, it extends forward as the angular
frequency is increasing. Along this branch, strong magnetic field of the vortex
destroys the superconductive phase in some circular domains of space around the symmetry axis.
On both branches the energy density of the $k$-chain solutions
is a set of $k$ tori located symmetrically with respect to the equatorial plane.

We constructed $k$-chains of the gauged spinning Q-balls we even number of constituents.
The building block of these solutions is the $\phi$-odd $n=1$ configuration whose angular dependence is given by the
spherical harmonic $Y^1_2(\theta,\varphi)$. This is a pair of spinning charged loops coupled to
a toroidal magnetic field, which forms a vortex encircling the configuration. By analogy with similar chains of boson
stars in the Einstein-Klein-Gordon theory \cite{Herdeiro:2020kvf}, one can look for solutions with odd number of constituents, they
would represent a deformation of the parity-even gauged spinning Q-balls considered recently in \cite{Loiko:2019gwk}.
However, we were not able to find such solutions, it is possible that the odd chains may exist for higher values of the
winding number $n$.

As another direction for future work, it would be interesting to study self-gravitating counterparts of these
negative parity chain solutions, they may give rise to a new type of gauged axially-symmetric
boson stars coupled to a solenoidal magnetic field.

\section*{Acknowledgements}
The work was supported by Ministry of Science and High Education of
Russian Federation, project FEWF-2020-0003. Computations were performed
on the cluster HybriLIT (Dubna).

 \begin{small}
 
 \end{small}


\begin{thebibliography}{99}
\bibitem{Rosen}G.~Rosen, J.\ Math.\ Phys.\ {\bf 9} (1968) 996, 999
\bibitem{Friedberg:1976me}
  R.~Friedberg, T.~D.~Lee and A.~Sirlin,
  Phys.\ Rev.\ D {\bf 13} (1976) 2739
\bibitem{Coleman:1985ki}
  S.~R.~Coleman,
  Nucl.\ Phys.\ B {\bf 262} (1985) 263
   Erratum: [Nucl.\ Phys.\ B {\bf 269} (1986) 744].
\bibitem{Lee:1991ax}
  T.~D.~Lee and Y.~Pang,
  Phys.\ Rept.\  {\bf 221} (1992) 251
 \bibitem{Shnir2018}
Y.M.~Shnir,
{\it 'Topological and Non-Topological Solitons in Scalar Field Theories'},
Cambridge University Press, 2018.
\bibitem{Radu:2008pp}E.~Radu and M.S.~Volkov,
Phys.\ Rept.\  {\bf 468}  (2008)  101.
\bibitem{Levin:2010gp}
  A.~Levin and V.~Rubakov,
  Mod.\ Phys.\ Lett.\ A {\bf 26} (2011) 409.
\bibitem{Loiko:2018mhb}
  V.~Loiko, I.~Perapechka and Y.~Shnir,
  Phys.\ Rev.\ D {\bf 98} (2018) no.4,  045018
\bibitem{Lee:1988ag}
K.~M.~Lee, J.~A.~Stein-Schabes, R.~Watkins and L.~M.~Widrow,
Phys. Rev. D \textbf{39} (1989), 1665
\bibitem{Lee:1991bn}
  C.~H.~Lee and S.~U.~Yoon,
  Mod.\ Phys.\ Lett.\ A {\bf 6} (1991) 1479.
\bibitem{Kusenko:1997vi}
  A.~Kusenko, M.~E.~Shaposhnikov and P.~G.~Tinyakov,
  Pisma Zh.\ Eksp.\ Teor.\ Fiz.\  {\bf 67} (1998) 229
   [JETP Lett.\  {\bf 67} (1998) 247]
\bibitem{Anagnostopoulos:2001dh}
  K.~N.~Anagnostopoulos, M.~Axenides, E.~G.~Floratos and N.~Tetradis,
  Phys.\ Rev.\ D {\bf 64} (2001) 125006
\bibitem{Gulamov:2015fya}
  I.~E.~Gulamov, E.~Y.~Nugaev, A.~G.~Panin and M.~N.~Smolyakov,
  Phys.\ Rev.\ D {\bf 92} (2015) no.4,  045011
\bibitem{Gulamov:2013cra}
  I.~E.~Gulamov, E.~Y.~Nugaev and M.~N.~Smolyakov,
  Phys.\ Rev.\ D {\bf 89} (2014) no.8,  085006
\bibitem{Panin:2016ooo}
  A.~G.~Panin and M.~N.~Smolyakov,
  Phys.\ Rev.\ D {\bf 95} (2017) no.6,  065006
\bibitem{Nugaev:2019vru}
E.~Y.~Nugaev and A.~V.~Shkerin,
J. Exp. Theor. Phys. \textbf{130} (2020) no.2, 301-320
\bibitem{Loginov:2020xoj}
A.~Y.~Loginov and V.~V.~Gauzshtein,
Phys. Rev. D \textbf{102} (2020) no.2, 025010
\bibitem{Loginov:2020lwg}
A.~Y.~Loginov and V.~V.~Gauzshtein,
[arXiv:2009.12818 [hep-th]].
 \bibitem{Witten:1984eb}
  E.~Witten,
  Nucl.\ Phys.\ B {\bf 249} (1985) 557.
\bibitem{Forgacs:2020vcy}
P.~Forg\'acs and \'A.~Luk\'acs,
Phys. Rev. D \textbf{102} (2020) no.7, 076017
\bibitem{Volkov:2002aj}M.S.~Volkov and E.~Wohnert,
Phys.\ Rev.\  D  {\bf 66} (2002)  085003.
\bibitem{Kleihaus:2005me}
  B.~Kleihaus, J.~Kunz and M.~List,
  Phys.\ Rev.\ D {\bf 72} (2005) 064002
\bibitem{Kleihaus:2007vk}
  B.~Kleihaus, J.~Kunz, M.~List and I.~Schaffer,
  Phys.\ Rev.\ D {\bf 77} (2008) 064025
\bibitem{Loiko:2019gwk}
V.~Loiko and Y.~Shnir,
Phys. Lett. B \textbf{797} (2019), 134810
\bibitem{Davis:1988jp}
  R.~L.~Davis and E.~P.~S.~Shellard,
  Phys.\ Lett.\ B {\bf 207} (1988) 404.
\bibitem{Davis:1988ij}
  R.~L.~Davis and E.~P.~S.~Shellard,
  Nucl.\ Phys.\ B {\bf 323} (1989) 209.
\bibitem{Garaud:2013iba}
  J.~Garaud, E.~Radu and M.~S.~Volkov,
  Phys.\ Rev.\ Lett.\  {\bf 111} (2013) 171602
\bibitem{Shiromizu:1998eh}
  T.~Shiromizu,
  Phys.\ Rev.\ D {\bf 58} (1998) 107301
\bibitem{Kleihaus:1999sx}
  B.~Kleihaus and J.~Kunz,
  Phys.\ Rev.\ D {\bf 61} (2000) 025003
\bibitem{Kleihaus:2000hx}
B.~Kleihaus and J.~Kunz,
Phys. Rev. Lett. \textbf{85} (2000), 2430-2433
\bibitem{Kleihaus:2003nj}
  B.~Kleihaus, J.~Kunz and Y.~Shnir,
  Phys.\ Lett.\ B {\bf 570} (2003) 237
 \bibitem{Kleihaus:2003xz}
  B.~Kleihaus, J.~Kunz and Y.~Shnir,
  Phys.\ Rev.\ D {\bf 68} (2003) 101701
\bibitem{Kleihaus:2004is}
  B.~Kleihaus, J.~Kunz and Y.~Shnir,
  Phys.\ Rev.\ D {\bf 70} (2004) 065010
\bibitem{Kleihaus:2004fh}
  B.~Kleihaus, J.~Kunz and Y.~Shnir,
  Phys.\ Rev.\ D {\bf 71} (2005) 024013
\bibitem{Teh:2004bq}
R.~Teh and K.~Wong,
J. Math. Phys. \textbf{46} (2005), 082301
\bibitem{Paturyan:2004ps}
V.~Paturyan, E.~Radu and D.~Tchrakian,
Phys. Lett. B \textbf{609} (2005), 360-366
 \bibitem{Kleihaus:2005fs}
  B.~Kleihaus, J.~Kunz and U.~Neemann,
  Phys.\ Lett.\ B {\bf 623} (2005) 171
\bibitem{Kunz:2006ex}
  J.~Kunz, U.~Neemann and Y.~Shnir,
  Phys.\ Lett.\ B {\bf 640} (2006) 57
  \bibitem{Kunz:2007jw}
  J.~Kunz, U.~Neemann and Y.~Shnir,
  Phys.\ Rev.\ D {\bf 75} (2007) 125008
\bibitem{Lim:2011ra}
K.~Lim, R.~Teh and K.~Wong,
J. Phys. G \textbf{39} (2012), 025002
\bibitem{Teh:2014zea}
R.~Teh, A.~Soltanian and K.~Wong,
Phys. Rev. D \textbf{89} (2014) no.4, 045018
\bibitem{Krusch:2004uf}
  S.~Krusch and P.~Sutcliffe,
  J.\ Phys.\ A {\bf 37} (2004) 9037
\bibitem{Shnir:2009ct}
  Y.~Shnir and D.~H.~Tchrakian,
  J.\ Phys.\ A {\bf 43} (2010) 025401
\bibitem{Shnir:2015aba}
  Y.~Shnir,
  Phys.\ Rev.\ D {\bf 92} (2015) no.8,  085039
\bibitem{Herdeiro:2020kvf}
C.~A.~R.~Herdeiro, J.~Kunz, I.~Perapechka, E.~Radu and Y.~Shnir,
[arXiv:2008.10608 [gr-qc]].\\
C.~A.~R.~Herdeiro, J.~Kunz, I.~Perapechka, E.~Radu and Y.~Shnir, work in progress.
\bibitem{pardiso}
N.I.M.~Gould,  J.A.~Scott and Y.~Hu,
ACM Transactions on Mathematical Software {\bf 33}~(2007)~10;\\
O.~Schenk and K.~G\"artner
Future Generation Computer Systems \textbf{20} (3) (2004) 475.
\bibitem{schoen}
 W. Sch\"onauer and R. Wei\ss ,
 J. Comput. Appl. Math. 27, 279 (1989) 279;
 \\
 M. Schauder, R. Wei\ss\ and W. Sch\"onauer,
 Universit\"at Karlsruhe, Interner Bericht Nr. 46/92 (1992).


 \end{thebibliography}
 \end{document}